\documentclass[preprint]{elsarticle}
\usepackage{lineno,hyperref,graphicx,amssymb,mathtools,comment,
color,multirow,gensymb,subcaption,epstopdf,threeparttable,siunitx,
xspace,hypcap}
\modulolinenumbers[5]

\DeclareSIUnit\sample{S}
\DeclareSIUnit\bits{bits}

\newcommand*\patchAmsMathEnvironmentForLineno[1]{%
  \expandafter\let\csname old#1\expandafter\endcsname\csname #1\endcsname
  \expandafter\let\csname oldend#1\expandafter\endcsname\csname end#1\endcsname
  \renewenvironment{#1}%
     {\linenomath\csname old#1\endcsname}%
     {\csname oldend#1\endcsname\endlinenomath}}%
\newcommand*\patchBothAmsMathEnvironmentsForLineno[1]{%
  \patchAmsMathEnvironmentForLineno{#1}%
  \patchAmsMathEnvironmentForLineno{#1*}}%
\AtBeginDocument{%
\patchBothAmsMathEnvironmentsForLineno{equation}%
\patchBothAmsMathEnvironmentsForLineno{align}%
\patchBothAmsMathEnvironmentsForLineno{flalign}%
\patchBothAmsMathEnvironmentsForLineno{alignat}%
\patchBothAmsMathEnvironmentsForLineno{gather}%
\patchBothAmsMathEnvironmentsForLineno{multline}%
}
\makeatletter
\renewcommand\tableofcontents{%
    \@starttoc{toc}%
}
\makeatother

\newif\ifshowchanges
\showchangesfalse

\ifshowchanges
\usepackage{soul}
\soulregister\cite7
\soulregister\ref7

\fi

\begin{document}

\begin{frontmatter}

\title{Study of Silicon Photomultiplier Radiation Hardness with Proton Beam from the JULIC Cyclotron}

\author[IKP]{T.~Tolba\corref{CorrAut}\fnref{UHH}}
\author[IKP]{D.~Grzonka\corref{CorrAut}}
\author[IKP]{T.~Sefzick}
\author[IKP]{J.~Ritman}

\address[IKP]{Institut f{\"u}r Kernphysik, Forschungszentrum J{\"u}lich, 52428 J{\"u}lich, Germany}
\cortext[CorrAut]{Corresponding author}
\fntext[UHH]{Present address: Institut f{\"u}r Experimentalphysik, University of Hamburg, 22761 Hamburg, Germany.}

\begin{abstract}
In this work we study the performance of silicon photomultiplier (SiPM) light sensors after exposure to the JULIC cyclotron proton beam of energy $\sim$~\SI{39}{\mega\electronvolt}, relative to their performance before exposure. The SiPM devices used in this study show a significant change in their behavior and a downward shift of their breakdown voltage by as much as $\sim$~0.4$\pm$0.1 V. Single photon measurements appear to be no longer possible for the SiPMs under study after exposure to a dose of $\sim$~0.2 Gy (corresponding to an integrated proton flux of $\sim$~$\phi_{p}$=1.06x10$^{8}$ p/cm$^{2}$). No visible damage to the surface of the devices was caused by the exposure.
\end{abstract}


\end{frontmatter}
\section{Introduction}
\label{sec:sec-intro}
Silicon photomultipliers (SiPMs) are multi-pixel semiconductor devices, with pixels (microcells) arranged on a common silicon substrate. Each microcell is a Geiger-mode avalanche photodiode (GM-APD), working above the breakdown voltage ($\it{U_\textrm{bd}}$), and it has a resistor for passive quenching of the breakdown. SiPMs are designed to have high gain (typically $\sim$~10$^{6}$), high photon detection efficiency ({PDE})~\cite{SiPM_2018}, excellent time resolution, and a wide range spectral response. They can be used to detect light signals at the single photon level. Compared with traditional photomultiplier tubes (PMTs) SiPMs are insensitive to external magnetic fields, more compact and do not require high operating voltage.  These features make SiPMs very attractive photosensors for experiments where excellent particle detection is a key parameter.

An example of an experiment that intends to use SiPMs as photosensors to detect fast scintillation light is the barrel time of flight (barrel-ToF) detector \cite{BarTofTDR} of the future $\overline{\textrm{P}}$ANDA spectrometer at FAIR in Darmstadt, Germany \cite{PANDA}. The barrel-ToF detector will be located at $\sim$~\SI{50}{\cm} radial distance to the beam axis. This location exposes the barrel-ToF to high radiation levels. It is thus imperative to avoid severe degradation of the SiPMs' performance due to this exposure. The estimated average equivalent neutron dose on the barrel-ToF detector is in the order of $\sim$~9.13$\times{10^{9}}$~$\it{n_{eq}}$(\SI{1}{\mega\electronvolt})/cm$^{2}$ a year \cite{BarTofTDR} (the equivalent neutron dose can be calculated by multiplying the proton flux, \textbf{$\phi_{P}$}, by the hardness factor, $\kappa$, of silicon, which depends on the proton energy and can be deduced from \cite{Kfactor}). This estimated neutron equivalent flux is based on assuming fused silica material. Because the exact internal structure and doping concentrations of the SiPMs are not disclosed by vendors, it is difficult to accurately estimate/simulate a priori the effect of the radiation damage on the SiPM structure. Thus, the radiation dependence, as a function of energy and flux, of such devices must be measured experimentally.

The expected damage effect from the radiation exposure can be categorized depending on the energy loss process due to the interaction between the impinging radiation and the SiPM tile \cite{GCasse}, as follows:
\begin{enumerate}
\item Surface damage due to the Ionizing Energy Loss (IEL) process, which is usually caused by photons and light charged particles, e.g. electrons. The surface damage of the SiPM tiles can cause:
\begin{enumerate}
\item Charge build-up on the surface-protection oxide layer of the SiPM.
\item Increase in the leakage current of the SiPM. 
\end{enumerate}
\item Bulk damage due to the Non-Ionizing Energy Loss (NIEL) process, which is usually caused by heavier particles, e.g. protons, neutrons and pions. The bulk damage of a SiPM can cause:
\begin{enumerate}
\item Crystal defects in the bulk of the Si lattice. This is usually generated by the heavy particles that penetrate the bulk of the SiPM die and cause a mono-/multi-displacement of the Si atoms (the Primary Knock-on Atoms (PKA)).
\item Change of effective doping concentration by producing acceptor-like defects which modify the depletion (breakdown) voltage. 
\item Increase of charge carrier trapping, which leads to a loss of charge (signal). 
\item Increase of the gain due to the change in the breakdown voltage.
\item Easier thermal excitement of electrons and holes that causes an increase of the leakage current, hence the dark current noise. 
\end{enumerate}
\end{enumerate}

Numerous experimental investigations have been conducted to study the influence of proton radiation exposure on the performance of the SiPMs \cite{ HEERING2008,MUSIENKO2009,BOHN2009,MATSUMURA2009,MUSIENKO2015, HEERING2016, Lacombe2019} (the effects of radiation damage in SiPMs due to other heavy particles, e.g. neutrons, and gammas can be found in \cite{GARUTTI2019,ng2019}). However, the results from the different studies are found to be not consistent with each other, especially on the operational conditions of the SiPMs after exposure to high and low fluxes. Table~$\ref{tab:pre_work}$, summarizes the previous work on SiPMs radiation hardness studies with protons, which mostly were conducted at energies lower than \SI{212}{\mega\electronvolt} \cite{ HEERING2008,MUSIENKO2009,BOHN2009,MATSUMURA2009,MUSIENKO2015, Lacombe2019}. There is only one measurement at very high energy, \SI{23}{\giga\electronvolt} \cite{HEERING2016}. Comparing different measurements in Table~$\ref{tab:pre_work}$ requires non-trivial assumptions and there is ambiguity in interpreting data, which further supports the need for the experimental measurements. 

A first step of these studies is the irradiation test of the SiPMs at low energy, in order to understand the performance of these devices at their fundamental working conditions, i.e. in the dark and at room temperature, which is covered in this paper. These studies will be continued with light conditions and with higher proton momenta up to about 3 GeV/c, an energy region more relevant for experiments like the $\overline{\textrm{P}}$ANDA experiment, for which no previous studies of SiPMs concerning radiation hardness were performed.

\begin{table}
  \begin{center}
  \caption{Summary of existing studies on SiPMs radiation hardness. For each study, the table shows the proton energy used in the study \textbf{E$_{p}$}, proton fluence (\textbf{$\phi_{p}$}), the \SI{1}{\mega\electronvolt} neutron equivalent fluence (\textbf{$\phi_{n-eq}$}) and the absorbed dose (\textbf{Dose}).}
    \label{tab:pre_work}
    \begin{threeparttable}
    
    \begin{tabular}{c|c|c|c|c|l}
      \textbf{Reference} & \textbf{E$_{p}$} & \textbf{$\phi_{p}$} & \textbf{$\phi_{n-eq}$} & \textbf{Dose} & \textbf{Main Results}
      \\
       & \textbf{[MeV]} & \textbf{[P/cm$^{2}$]} & \textbf{[n$_{eq}$/cm$^{2}$]} & \textbf{[Gy]} & 
       \\
      \hline
      Heering & 212 & 3.00x10$^{13}$ & 8.00x10$^{12}$ & 1.68x10$^{4}$ & - Increase in leakage current 
      \\
       \cite{HEERING2008} (2008) & & & & & - Increase in dark count rate
      \\
      & & & & & - Decrease in gain
      \\
      & & & & & - At max. \textbf{$\phi_{p}$} SiPM are not working
      \\
      & & & & & 
      \\
      Musienko & 83 & 1.00x10$^{10}$ & 2.00x10$^{10}$ & 1.05x10$^{1}$ & - Increase in leakage current  
      \\
      \cite{MUSIENKO2009} (2009) & & & & & - Increase in dark count rate
      \\
      & & & & & - No change in $\it{U_\textrm{bd}}$ and R$_{q}$
      \\
      & & & & & - Reduction in PDE ($\textless$ \SI{10}{\percent})
      \\
      & & & & & - Reduction in gain ($\textgreater$ \SI{10}{\percent})
      \\
      & & & & & 
      \\
      Bohn & 212 & 3.00x10$^{10}$ & 2.40x10$^{10}$ & 1.68x10$^{1}$ & - Increase in leakage current  
      \\
      \cite{BOHN2009} (2009) & & & & & - Increase in dark noise
      \\
      & & & & & - At max. \textbf{$\phi_{P}$} SiPMs are working
      \\
      & & & & & - Reduction in PDE (\SI{4}{\percent} - \SI{49}{\percent})
      \\
      & & & & & 
      \\
      Matsumura & 53.3 & 2.80x10$^{10}$ & 4.80x10$^{10}$ & 4.20x10$^{1}$ & - Increase in leakage current 
      \\
     \cite{MATSUMURA2009} (2009) & & & & & - No significant change in gain
      \\
      & & & & & - At 21 Gy no photon counting
      \\
      & & & & & - Pulse height reduced at 42 Gy
      \\
      & & & & & 
      \\
      Heering & 62 & 6.00x10$^{12}$ & 1.20x10$^{13}$ & 8.02x10$^{3}$ & - Increase in dark current  
      \\
      \cite{HEERING2016} (2014) & & & & & - \SI{200}{\milli\volt} shift in $\it{U_\textrm{bd}}$
      \\
      & & & & & 
      \\
      Musienko & 62 & 1.00x10$^{12}$ & 2.00x10$^{12}$ & 1.34x10$^{3}$ & - Increase in leakage current 
      \\
      \cite{MUSIENKO2015} (2015) & & & & & - Increase in dark noise
      \\
      & & & & & - \SI{178}{\milli\volt} shift in $\it{U_\textrm{bd}}$
      \\
      & & & & & - Reduction in gain ($\textless$ \SI{38}{\percent})
      \\
      & & & & & 
      \\
      Heering & 23000 & 1.30x10$^{14}$ & 2.20x10$^{14}$ & & - Increase in dark noise
      \\
      \cite{HEERING2016} (2016) & & & & & - $\sim$~\SI{4}{\volt} shift in $\it{U_\textrm{bd}}$ 
      \\
      & & & & & - Reduction in PDE by \SI{25}{\percent} 
      \\
      & & & & & - At max. \textbf{$\phi_{p}$} SiPMs are working
      \\
      & & & & &
      \\
      Lacombe & 10 & 7x10$^{10}$ & 3.19x10$^{11}$ & 3.87x10$^{2}$ & - Increase in dark current  
      \\
      \cite{Lacombe2019} (2019) & 49.7 & 5x10$^{10}$ & 8.42x10$^{10}$ & 3.19x10$^{1}$ \tnote{*} & - No change in $\it{U_\textrm{bd}}$
      \\
      \hline
    \end{tabular}
\begin{tablenotes}
   \scriptsize \item[*]According to our calculation that depends on deducing the stopping power value of protons, with energy stated in the reference, in Si lattice, from the NIST database, and using the stated integrated flux, the dose value of this measurement must be $\sim$~7.88x10$^{1}$ Gy.  
  \end{tablenotes}
 \end{threeparttable}
 \end{center}
 \end{table}

\section{Instrumentation}
\label{sec:Inst}

\subsection{The Test Setup}
\label{sec:test_station} 
\begin{figure}
    \centering
    \includegraphics[width=0.8\linewidth]{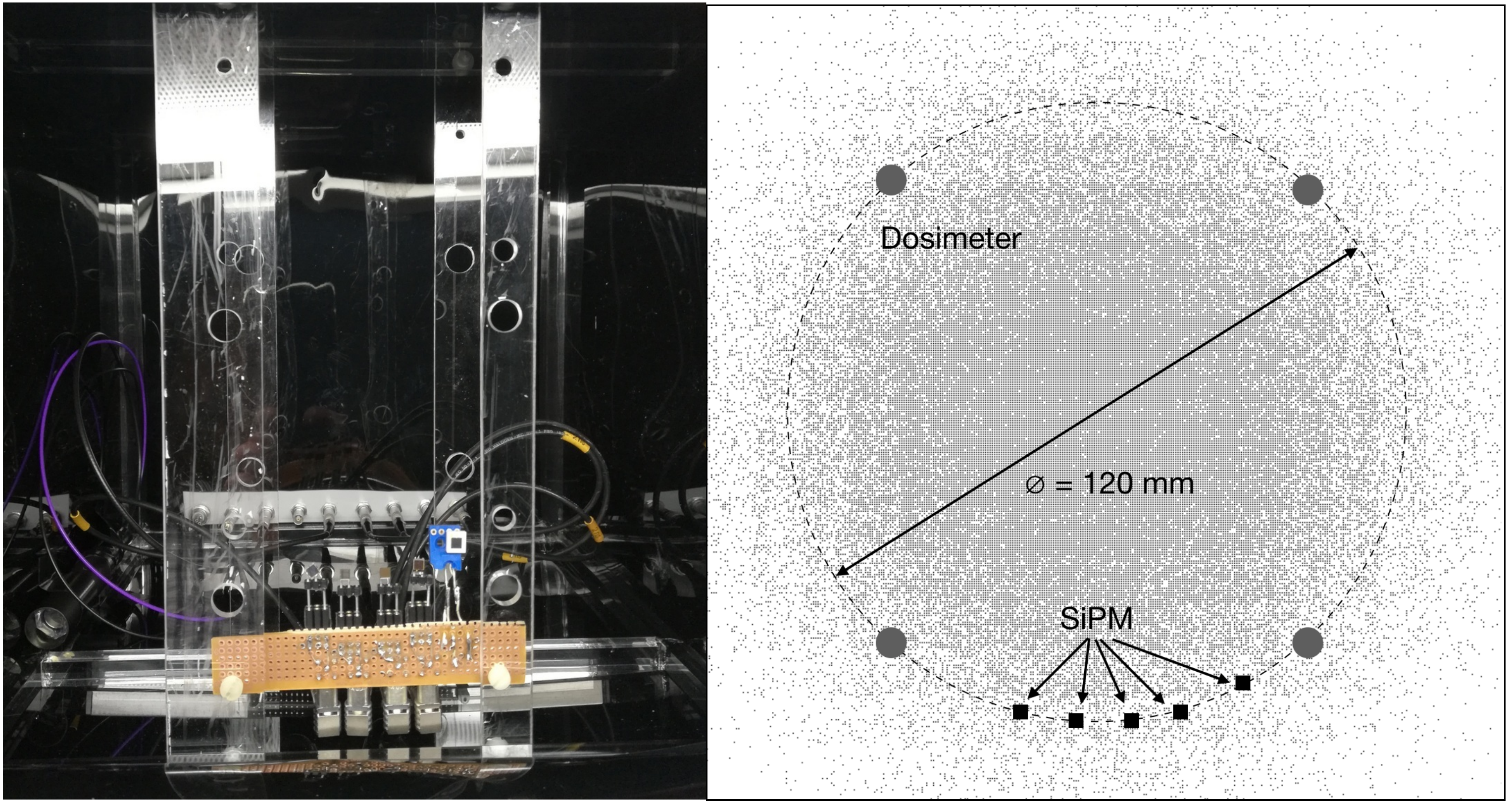}
    \caption{ Arrangement of the SiPMs in the irradiation box (left), positioned at \SI{6}{\cm} from the center of the beam axis, with four dosimeters as shown in the sketch on the (right). In the photo on the left side the dosimeters were not mounted.}
    \label{fig:CylBeam}
\end{figure}

Figure~$\ref{fig:CylBeam}$ shows a picture (left) with a sketch (right) of the radiation test station. The proton beam was produced by the JULIC cyclotron \cite{JULIC} as a H$^{-}$-beam of kinetic energy \SI{45}{\mega\electronvolt}, that is stripped in the exit foil. The setup was placed in air about \SI{2}{\meter} downstream of the exit window of the beam line. Energy loss and straggling in the window material and air resulted in a defocusing of the beam, which was further expanded by additional absorber material. The beam covers an area of about \SI{20}{\cm} diameter at the position of the SiPMs, thereby lowering the particle flux density. The beam energy was determined by a Bragg-peak measurement. A Plexiglas block, which included a radiation sensitive foil in the beam direction was mounted at the position of the SiPMs under the same beam conditions used for the SiPM measurement. The foil was darkened due to the radiation and shows the Bragg-curve as an intensity variation, which allows to determine the depth of the Bragg-peak and the corresponding beam energy. The beam energy seen by the SiPMs was found to be 39$\pm$1 MeV. The SiPMs were mounted in a closed light-tight box at a radius of \SI{6}{\cm} from the beam center in order to reduce the dose variation due to position uncertainty. The arrangement is sketched in Figure~$\ref{fig:CylBeam}$ (right). The beam has a Gaussian profile as a function of the distance from the axis. For the dose measurement, 4 calibrated Farmer chamber dosimeters  \cite{Dosimeters} were installed in addition at the same distance from the beam axis. Before installing the SiPMs, the beam was centered by adjusting the dose rates on the dosimeters to a comparable level resulting in an equal dose seen by the SiPMs. The sensitive volume of a dosimeter is \SI{0.6}{\cm}$^{3}$ with a sensitivity of 20 nC/Gy. Every second the accumulated charge is measured in each dosimeter in order to monitor the beam position and intensity during the irradiation. The area covered by a dosimeter was comparable to the area covered by a SiPM, resulting in a comparable mean dose rate. The absolute dose measurement error was below 10$\%$. The stability of the temperature inside the irradiation box was monitored by a DALLAS DS18B20 programmable resolution 1-wire digital temperature sensor controlled with a micro controller board "Arduino UNO control board" \cite{arduino}.

\subsection{The Photo-Sensors}
\label{sec:SiPMs}
In this study five SiPM devices of interest to the $\overline{\textrm{P}}$ANDA barrel-ToF detector were tested: two from KETEK~\cite{ketek} (PM3315-WB-BO and PM3325-WB-BO), one from Hamamatsu~\cite{hamamatsu} (S13360-3050CS), SensL~\cite{SensL} (MicroFC-30035-SMT) and AdvanSiD~\cite{AdvanSiD} (ASD-NUV3S-P-40). Detailed information for each device is given in Table~$\ref{tab:SiPM_prope}$.\\
The contacting of the SiPMs was done by soldering pins at the anode and cathode, which were connected to \SI{50}{\ohm} coax cables with LEMO plugs. The operating bias of the SiPMs (and the preamplifier) was supplied by a TTI QL564T power supply \cite{powersupply}.

\begin{table}
  \begin{center}
    \caption{Characteristics of the SiPM devices used in this study.}
    \label{tab:SiPM_prope}
    \begin{tabular}{c|c|c|c|c|c}
      \textbf{Device} & \textbf{Dimensions} & \textbf{Fill} & \textbf{Breakdown} & \textbf{Over} & \textbf{Microcell} \\
       & & \textbf{factor} & \textbf{voltage ($\it{U_\textrm{bd}}$)} & \textbf{voltage} & \textbf{size} \\
       &  \textbf{[mm$^{2}$]} & \textbf{[\% ]} & \textbf{@22 $\degree$C} [V] & \textbf{($\it{U_\textrm{ov}}$)} [V] & [$\mu$m] \\
      \hline
      KETEK-\SI{15}{\micro\meter} & 3 x 3 & 65 & 27.6$\pm$0.1 & 3.9 & 15 \\
      KETEK-\SI{25}{\micro\meter} & 3 x 3 & 65 & 25$\pm$0.2 & 5.7 & 25 \\
      Hamamatsu & 3 x 3 & 74 & 52$\pm$0.2 & 6.9 & 50 \\
      SensL & 3 x 3 & 64 & 24.8$\pm$0.2 & 5.0 & 35 \\
      AdvanSiD & 3 x 3 & 60 & 26.8$\pm$0.2 & 5.5 & 40 \\
      \hline
    \end{tabular}
  \end{center}
\end{table}
     
\subsection{The Data Acquisition (DAQ) System and Data Collection}
\label{sec:DAQ}
\begin{figure}
    \centering
   \includegraphics[width=0.6\linewidth]{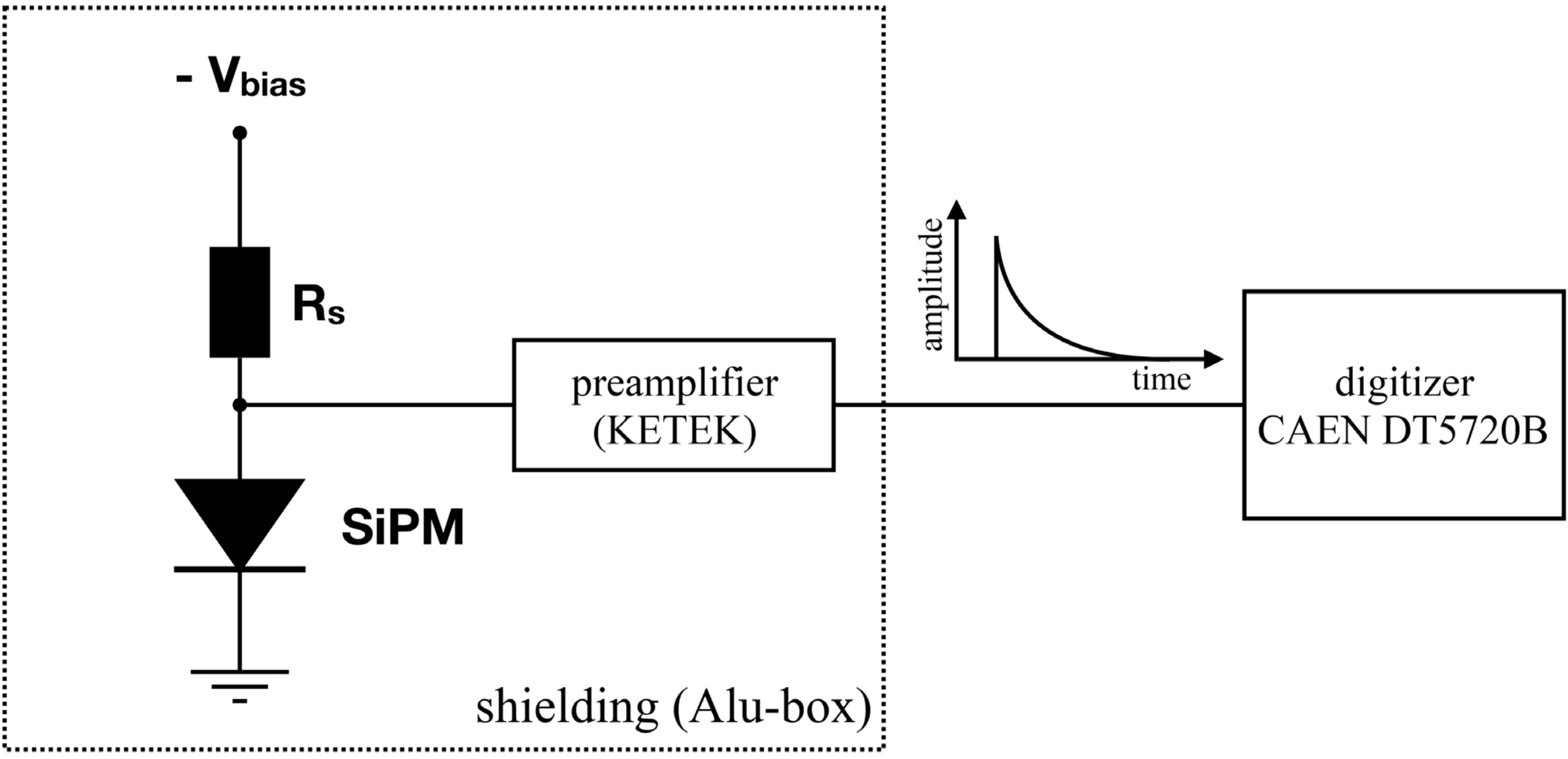}
    \caption{Schematic of the readout electronics. The SiPM and preamplifier were placed in an aluminum-box (indicated by the dotted line) for shielding.} 
    \label{fig:SchDia}
\end{figure}

The electronic circuit for the readout is shown in Figure~$\ref{fig:SchDia}$. The SiPMs were operated at 21 $\pm$ 0.2~\SI{}{\celsius} and $\it{U_{ov}}$ for each SiPM as listed in Table~$\ref{tab:SiPM_prope}$. The SiPM output signals were passed to a KETEK PEVAL-KIT MCX preamplifier unit \cite{KeteckPreamp}, resulting in a signal height amplification in the range of \numrange[range-phrase = --]{5}{10}~\SI{}{\milli\volt} for a single photon. The preamplified signals from the SiPMs were digitized by a $4-$channel CAEN DT5720B digitizer unit~\cite{caen} with 12 bit resolution, \SI{2}{\volt} dynamic range, and a maximum sampling rate of \SI{250}{\mega\hertz}. The captured pulses were sent to a PC and recorded by the DAQ CAEN Multi-PArameter Spectroscopy Software (CoMPASS)~\cite{CoMPASS} for further off-line processing with ROOT \cite{ROOT}.

Before irradiation, each SiPM was separately tested for its performance. In order to reduce the noise level, the irradiation box and the preamplifier were placed in an aluminum-box that was equipped with feedthroughs for power supplies and signal. In this configuration a noise level of the amplified signal of $\sim$~\SI{2}{\milli\volt} could be achieved, which was sufficiently low for a clear separation of the single photon signals from the background. Therefore the signals from the SiPMs were acquired by setting a \numrange[range-phrase = --]{4}{5}~mV threshold on the SiPM output signal amplitude in the discriminator node of the CoMPASS program. The digitizer acquisition time window was set to \SI{4}{\micro\second}, with a sampling rate of 250$\times{10^{6}}$ samples per second the waveforms were integrated over a time window of \SI{1}{\micro\second}.

\section{Data Analysis}
\label{sec:Data analsis}
\begin{figure}
    \centering    
 \includegraphics[width=0.7\linewidth]{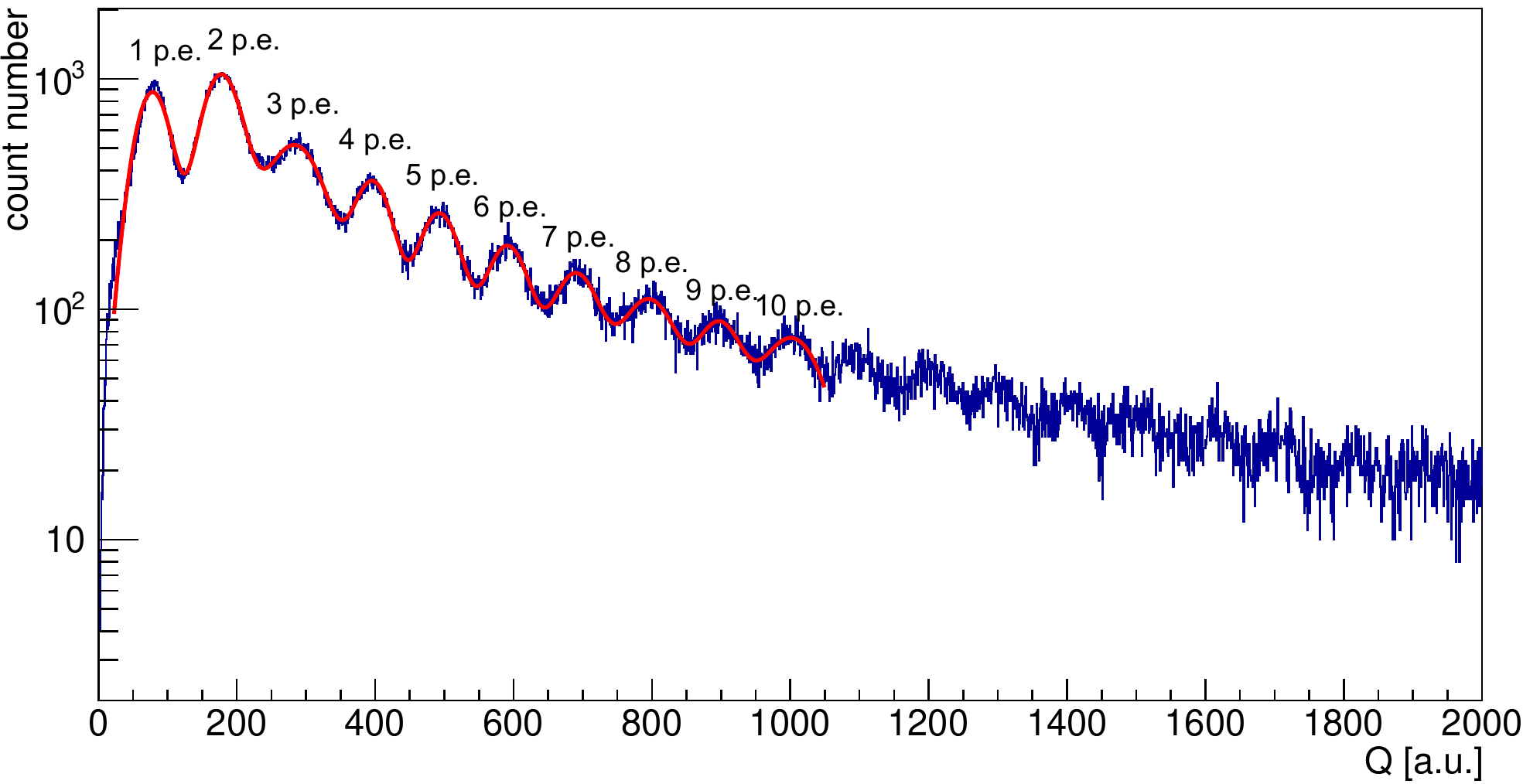}
 \caption{An example of the output charge spectrum of the AdvanSiD SiPM, taken in the dark, before exposure. The multi p.e.\ peaks have been fit with a sum of independent Gaussian functions (red line).}
\label{fig:Single_photon}
\end{figure}

The analysis strategy used in this experiment was similar to that explained in \cite{SiPM_2018}. The offline analysis first discriminates light signals from noise using a pulse finding algorithm (PFA), then calculates the total charge, $\it{Q}_{tot}$, collected by the SiPMs, for each event. The PFA selects signal pulses using a series of cuts. It first sets a lower limit on the pulse amplitude above the baseline. It then looks at the correlation between the pulse width and the corresponding integrated charge, as a 2D histogram. A ROOT graphical cut is used in this 2D plot to select and exclude the false signals and noise pulses with low charge and/or small width. The signal selection and noise rejection efficiency of the cuts are measured from the individual spectrum of each cut. The total charge for each identified signal is calculated by integrating the total ADC values in the pulse after baseline subtraction. The baseline is defined as the average of the waveform in the \SI{19}{\nano\second} time window prior to the trigger. Figure~$\ref{fig:Single_photon}$ shows an example of the output charge spectrum for the AdvanSiD SiPM, taken in the dark (before radiation exposure), where each peak above zero corresponds to a quantized number of photoelectrons, p.e.. The multi p.e. peaks were fit with the sum of independent Gaussian functions to estimate the gain.

\begin{figure}
    \centering    
 \includegraphics[width=1.\linewidth]{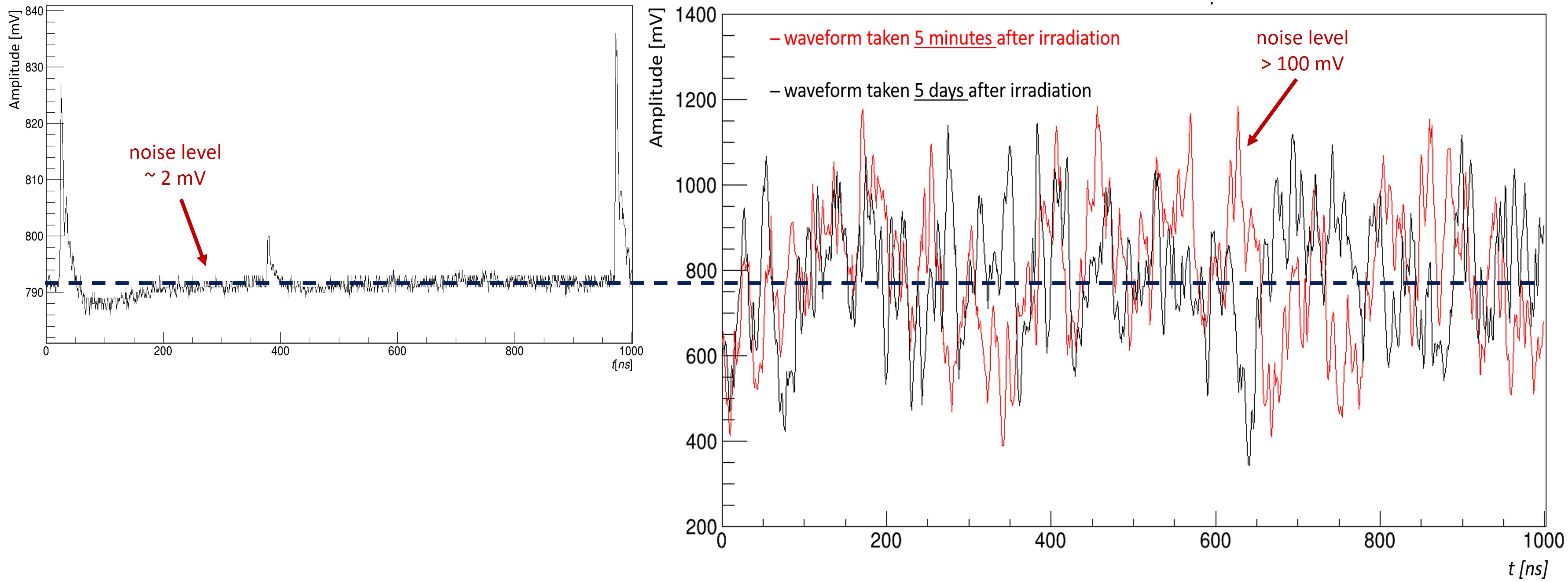}
 \caption{Typical signals of the SPiMs before (left side) and after (right side) irradiation with an integrated dose of 10 Gy. The level as well as the width of the noise band is drastically increased.}
 \label{fig:sipmsignal}
\end{figure}

For the radiation hardness measurements the irradiation box was placed at the beam line but outside of the aluminum box, which led to a drastic increase in the noise level. Therefore, it was planned to conduct the irradiation test in several steps, with measurements in between. In these measurements, the SiPMs were not under bias during irradiation. In view of the information available about the acceptable dose rates from the existing measurements, we planned for five steps of irradiation with a total dose of $\sim$~50 Gy (with an instantaneous proton flux of $\phi_{P}$=1.48x10$^{8}$ p/cm$^{2}\cdot$s for \SI{180}{\second}, corresponding to an integrated proton flux of $\phi_{P}$=2.67x10$^{10}$ p/cm$^{2}$ and neutron equivalent fluence of $\phi_{n-eq}$=5.07x10$^{10}$ n$_{eq}$/cm$^{2}$) in each step. But already after the first irradiation step all SiPMs used in this study seemed to be completely insensitive to resolve single photoelectrons. The noise increased to a level of about \SI{100}{\milli\volt} compared to the \SI{2}{\milli\volt} before irradiation, as shown in Fig. \ref{fig:sipmsignal} (right). The  typical signals of the SiPMs before irradiation were well separated and far above background with a reasonable signal rate (Fig. \ref{fig:sipmsignal} (left)). The signal structure after irradiation confirms the drastic increase of the dark current with a drastically increased dark signal rate (Fig. \ref{fig:sipmsignal} (right)). The data shown in this figure were collected \SI{5}{\min} (red waveform) and once again 5~days (black waveform) after irradiation of the SiPMs (the SiPMs were always stored in a fume-hood under normal atmosphere at controlled temperature of $\sim$~\SI{21}{\celsius}). In view of a signal height of $\sim$~\SI{20}{\milli\volt} for a single photon, the devices are not usable for sensitive measurements at the few photoelectrons level. In view of these initial results, and in order to get an insight to the device's performance, we analyzed the critical parameters of the SiPMs.

\subsection{Current-Voltage (I-V) Curve Studies}
\label{sec:IV_Plots}
\begin{figure}
   \centering
   \includegraphics[width=0.45\linewidth]{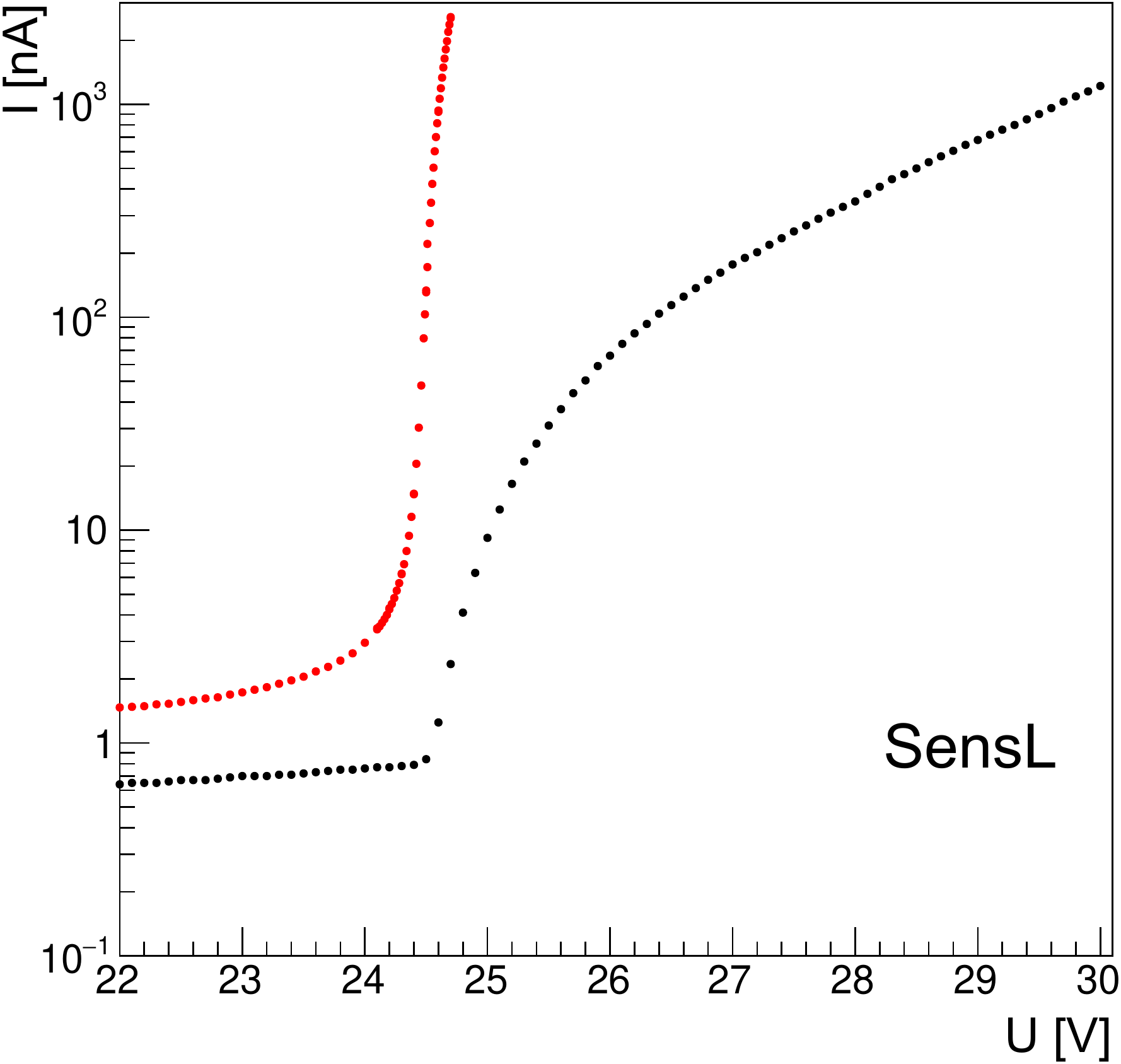} ~~\includegraphics[width=0.45\linewidth]{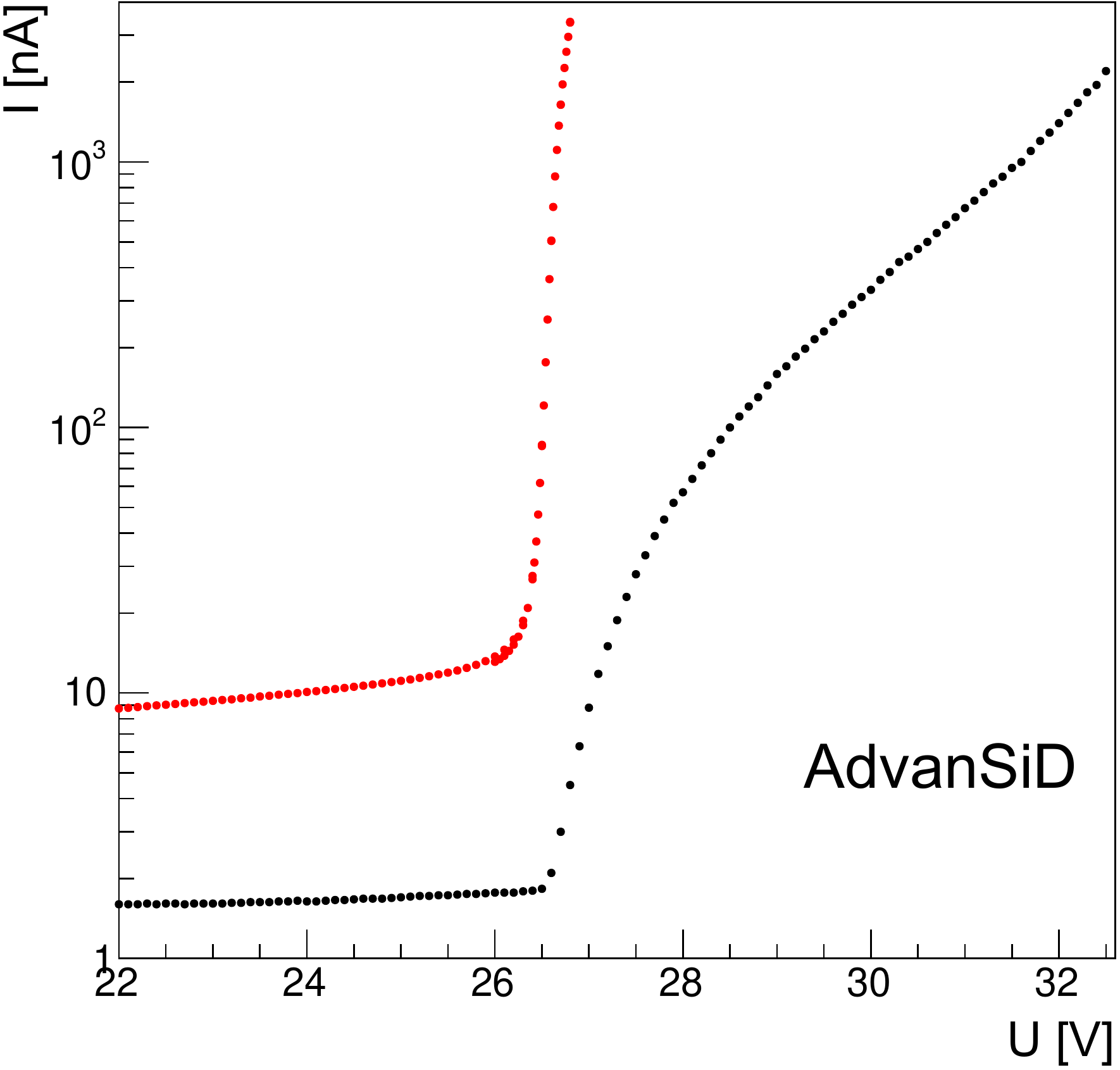}
   
   \vspace{1.5mm}
   
   \includegraphics[width=0.45\linewidth]{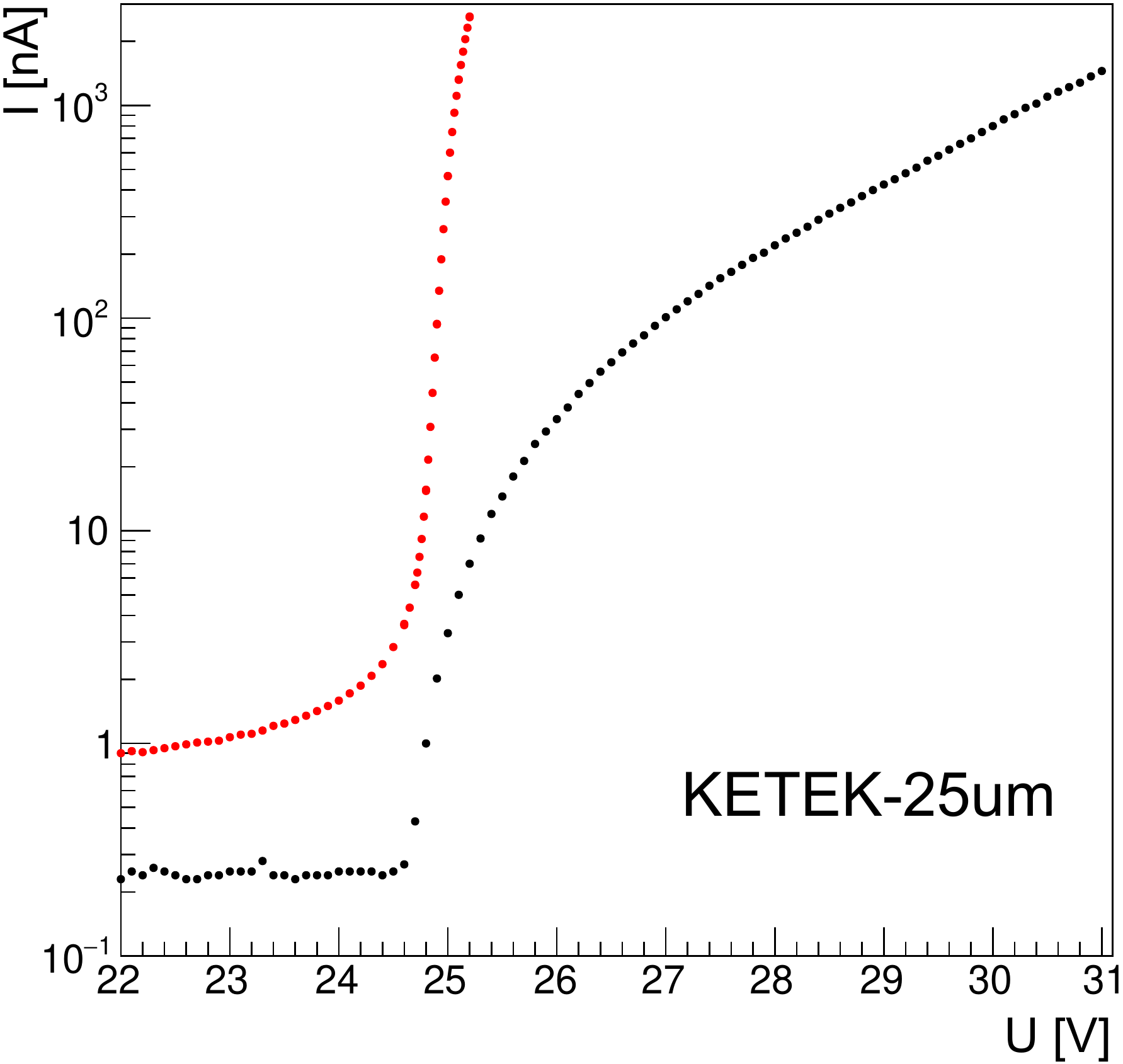} ~~\includegraphics[width=0.45\linewidth]{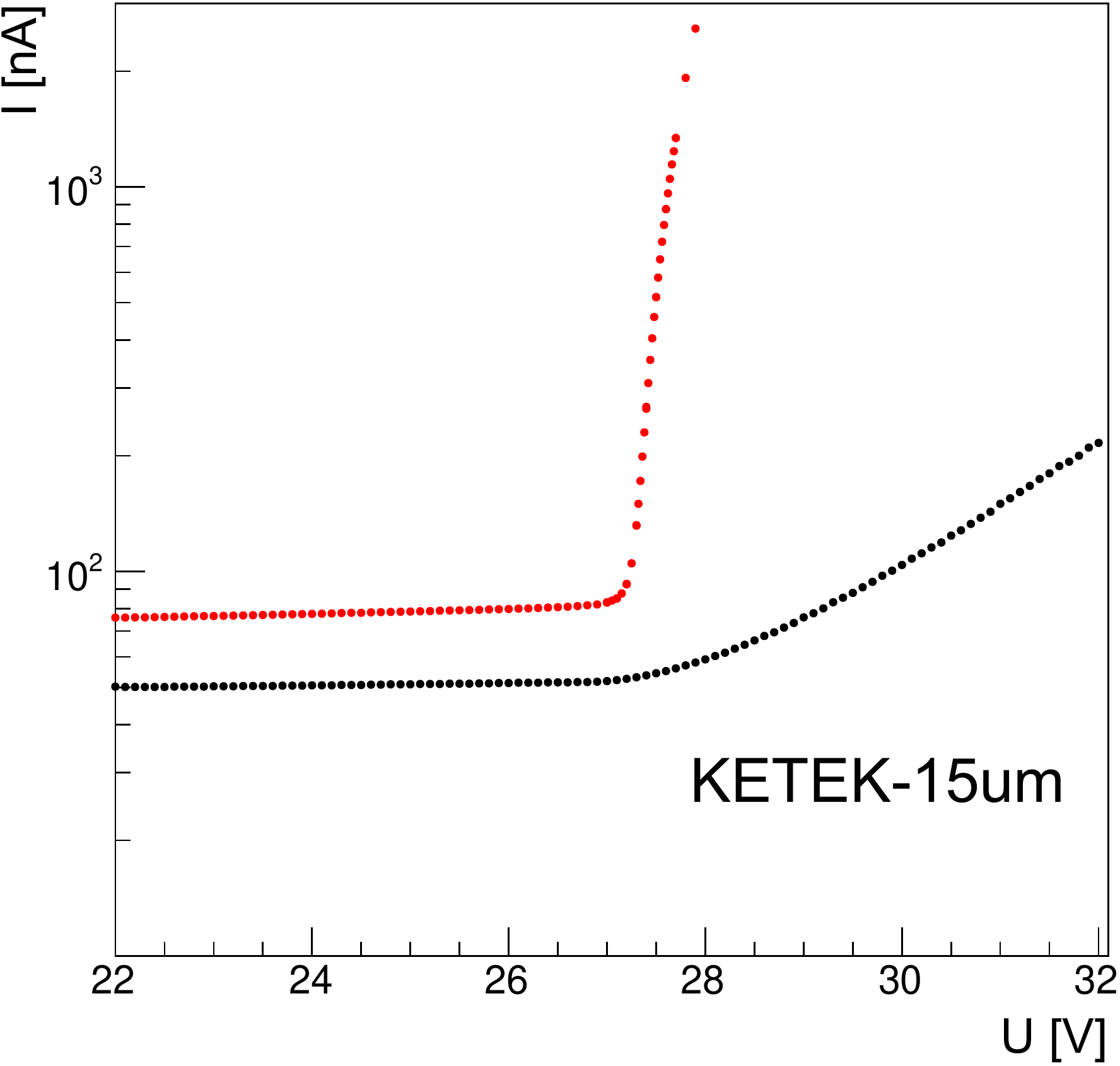}
   
   \vspace{1.5mm}
   
   \includegraphics[width=0.45\linewidth]{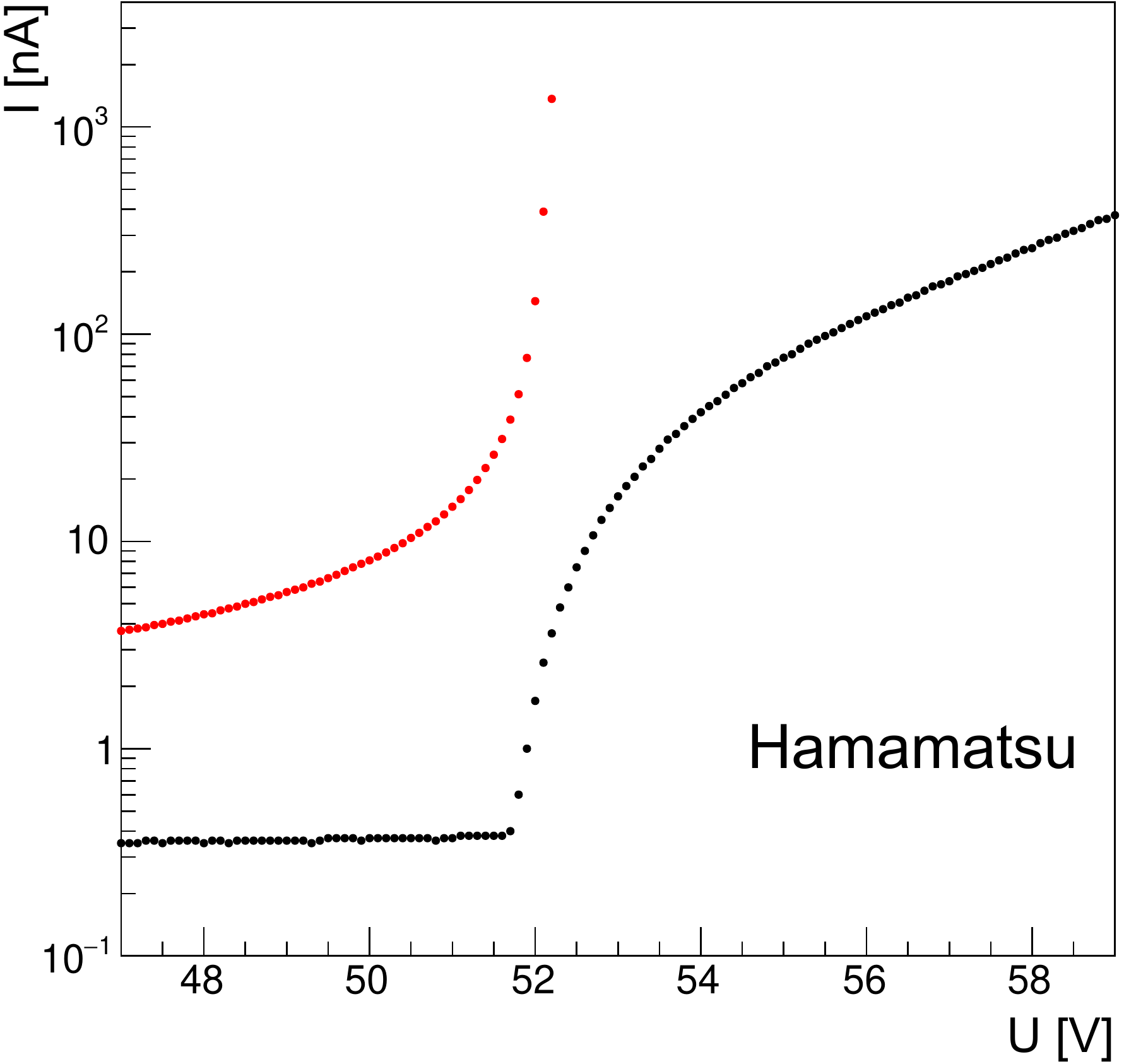} 
   \caption{I-V curves before (black symbols) and after (red symbols) irradiation for SensL (top-left), AdvanSiD (top-right), KETEK-\SI{25}{\micro\meter} (middle-left), KETEK-\SI{15}{\micro\meter} (middle-right) and Hamamatsu (bottom) SiPMs.}
   \label{fig:IV_Curves}
\end{figure}

Because the SiPMs I-V curve can reveal subtle changes in their characteristics, we measured the I-V behaviour of each device before and after irradiation. All the measurements were taken in the dark. In this measurement, each SiPM was connected to a picoammeter (Keithley 6485~\cite{picoammeter}) and its leakage current as a function of the bias voltage was measured. The bias voltage was incremented in steps of \SI{0.5}{\volt} up to $\it{U_\textrm{bias}}$~=\SI{20}{\volt}, except for the Hamamatsu; up to $\it{U_\textrm{bias}}$~=\SI{40}{\volt}, after which the step size was reduced to \SI{0.1}{\volt}, to improve the accuracy of the breakdown voltage determination. For some of the irradiated devices, the step size was further reduced to \SI{0.02}{\volt} at the area where strong change in current was observed, up to a current increase of about \SI{1000}{\nano\ampere}. The I-V curves were studied in two biasing modes; ramp-up and ramp-down. No changes in the I-V data were found between the two ways of measurement. Another method; the inverse logarithmic derivative (ILD) \cite{Garutti2016}, was also used to determine the $\it{U_\textrm{bd}}$, however the results were not consistent with that observed from the collected I-V data. One possible reason for this inconsistency is that the ILD method is better applied for measurements with light, while these measurements were taken in dark. However, the $\it{U_\textrm{bd}}$ values presented here were estimated from the I-V data and the difference in the values between the two methods were added to the uncertainty estimation of the $\it{U_\textrm{bd}}$. The effective resolution of the system was dominated by noise pickup, which was on the order of \SI{100}{\pico\ampere}. Figure~$\ref{fig:IV_Curves}$ shows the I-V curves for the five types of SiPMs. The onset of breakdown is clearly visible in all SiPMs before irradiation (black markers)\footnote{Due to an error in the first I-V curve measurements before irradiation, the I-V curve before irradiation of the Hamamatsu SiPM presented here was conducted on a SiPM device delivered at later time, i.e. from different production batch.}. After irradiation (red markers) the behavior drastically changed. The avalanche effect, i.e. a strong current increase at a certain voltage,  is still visible but the breakdown voltage is about $\sim$~0.4$\pm$0.1 V lower for all SiPMs compared to the $\it{U_\textrm{bd}}$ mean value before irradiation. Furthermore the slope of the dark current increased by approximately 2 orders of magnitude.

\subsection{SiPM Performance with Light}
\begin{figure}
   \centering
   \includegraphics[width=1.\linewidth]{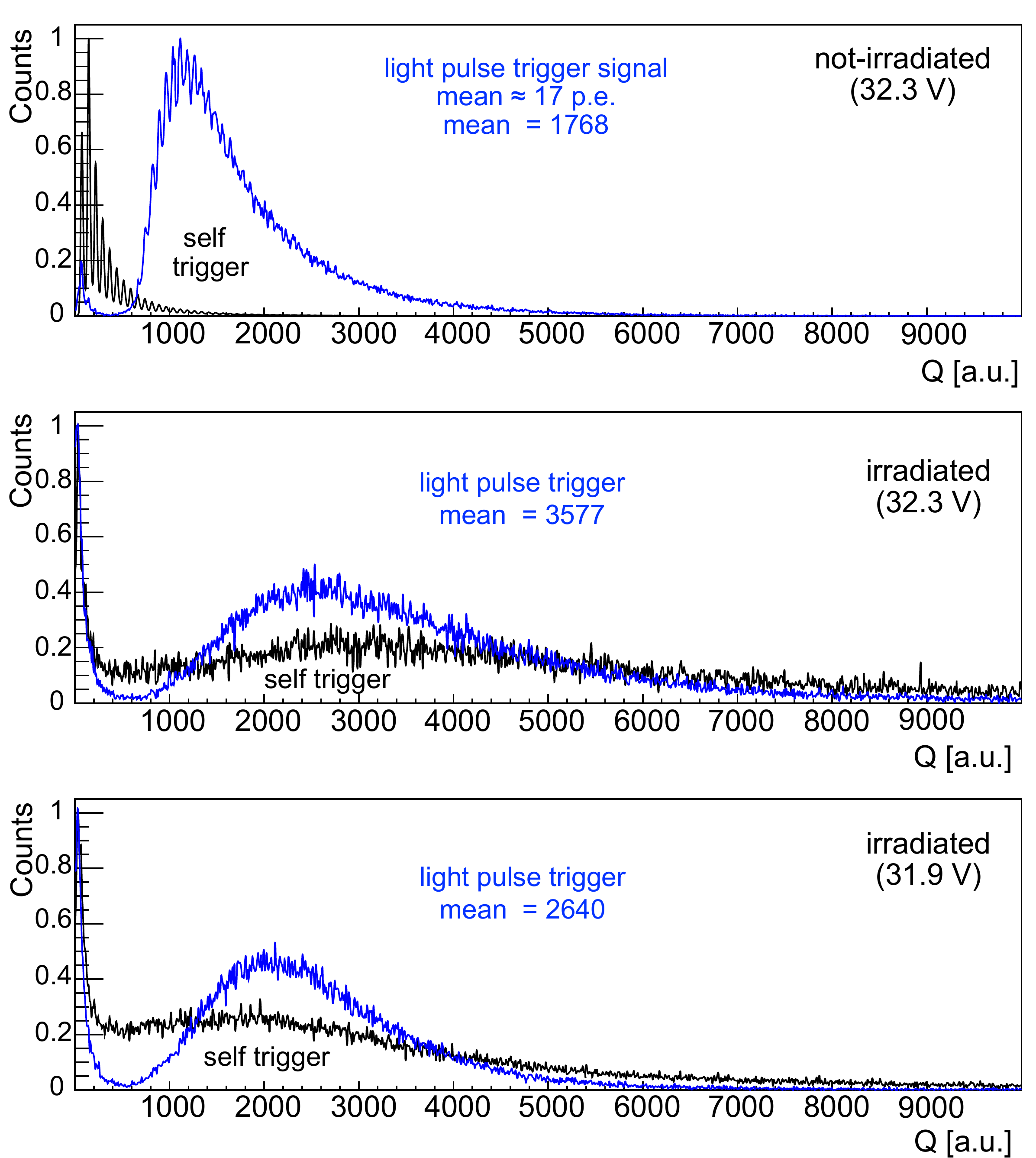}    
    \caption{Comparison of a new "not-irradiated" AdvanSid SiPM (upper plot) and an irradiated one, with 1 Gy integration dose (middle and lower plots). The SiPMs were illuminated by the same light pulses with a mean of about 17 photoelectrons. The black curve results from selftriggering, while the signals of the blue curves were triggered by the light pulse.}
      \label{fig:sipmled}
\end{figure}

A test of the irradiated SiPMs sensitivity to photoelectrons was performed using light pulses from a laser diode. For this measurement, two SiPMs from AdvanSiD were used; an irradiated one (with an integrated dose of 1 Gy) and a new one of the same type. Both devices were illuminated by a pulsed blue laser diode and operated at the same bias voltage. In addition, another measurement of the irradiated device at a reduced bias voltage that corresponds to the reduction in $\it{U_\textrm{bd}}$ after exposure, was conducted in order to study the effect of the reduced operating voltage on the excess rate of the dark noise. The measurements were done by placing the SiPMs alternately at the same position while keeping the light system unchanged to achieve the same illumination. The output signals from both SiPMs were captured and the total charges were compared. Figure~$\ref{fig:sipmled}$, shows the corresponding output charge spectrum for both devices. The individual distributions were normalized by setting the maxima in the charge distributions to 1. The irradiated SiPM (middle and lower plots) remained sensitive to the light source at the high p.e.\ level. However, multi-p.e. signals are not separable any more. The increase in the gain of the irradiated SiPM (middle plot), compared to that with no irradiation (upper plot), is due to the fact that both devices were operated at the same operating voltage. This can be explained as following; due to the reduction in the $\it{U_\textrm{bd}}$ value of the irradiated SiPMs, as explained in section~$\ref{sec:IV_Plots}$, operating the two devices at the same operating voltage results in an increase in the $\it{U_\textrm{ov}}$, hence resulted in increasing the gain of this device. The lower plot shows the charge distribution for the measurement at \SI{0.4}{\volt} lower operating voltage (with respect to the operating bias of the non-irradiated SiPM). While the signal amplitudes show a strong reduction (i.e. reduced gain), the device performance is comparable to its measurement at the higher operating voltage.  The noise signal amplitudes are reduced similarly and the high background level compared to the light signals remains. 

\label{sec:Gain_dark}

\subsection{SiPMs Visual Inspection}
\label{sec:vis_ins}
A visual inspection of the irradiated SiPMs was carried out at the end of the tests using an optical microscope, with magnification power ranging from 20x to 128x. The surfaces of the devices were carefully inspected, and photographs of specific locations were taken, before and after the irradiation tests. No visible evidence for damage was found on the outer surface of the SiPMs or at the microcell level.

\section{Data Taking During Irradiation}
\label{sec:DTDI}
In order to follow the radiation damage as a function of the dose rate, an additional measurement was performed with one SiPM (the AdvanSiD) during irradiation. The gain, dark current rate, prompt cross-talk and correlated noise probabilities were studied at different radiation doses and compared to their values in the absence of radiation exposure. To reduce the noise level, the irradiation box that houses the SiPMs was shielded by a layer of aluminum foil, while the preamplifier was placed at a distance of about \SI{2}{\meter} from the irradiation box, in a separate metal box. With these measures a noise level of about \SI{3}{\milli\volt} was achieved. Furthermore, the cyclotron beam current was reduced to a dose rate of 0.001 Gy/s.

\begin{figure}
   \centering
   \includegraphics[width=0.48\linewidth]{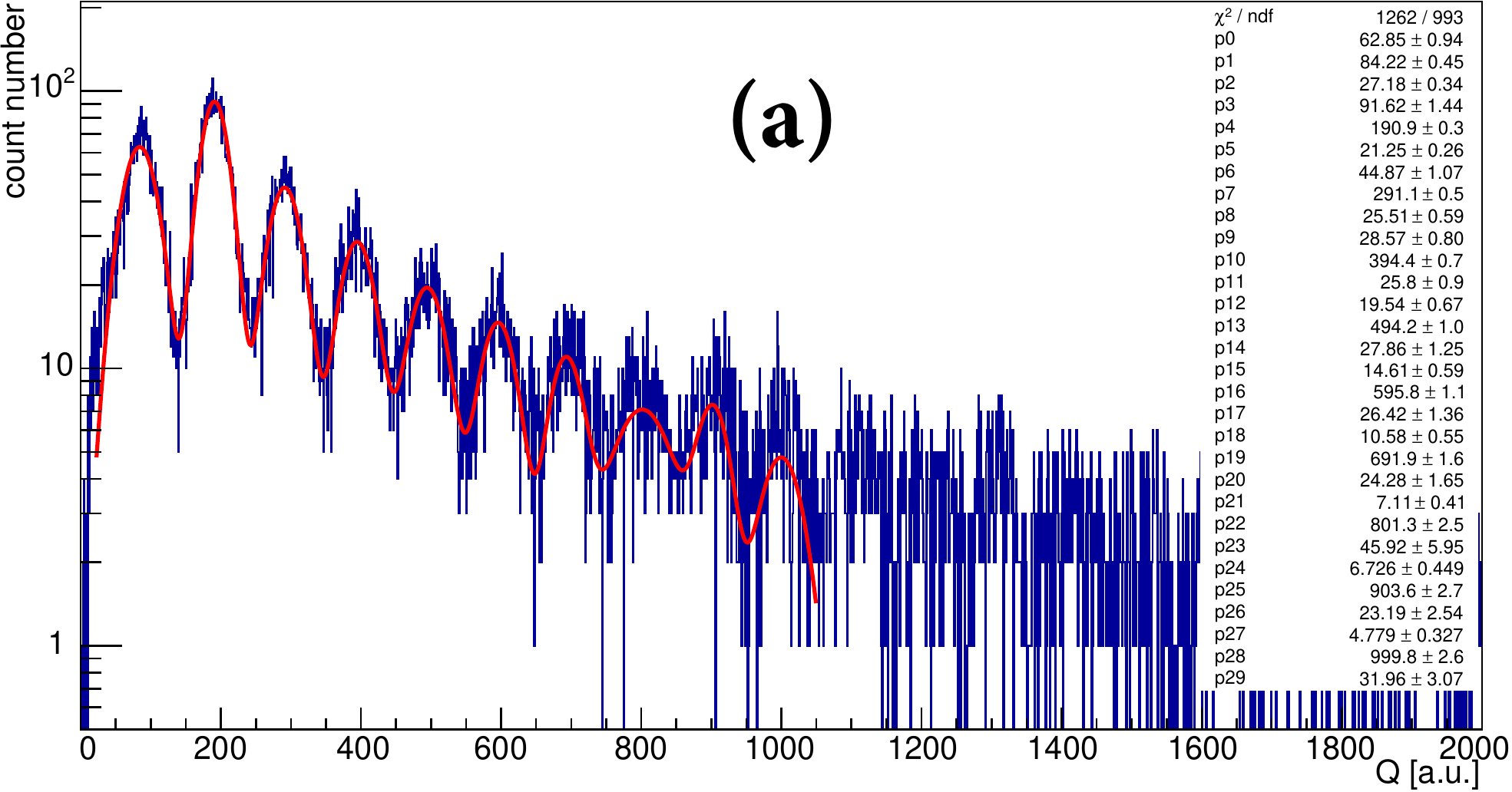} ~~\includegraphics[width=0.48\linewidth]{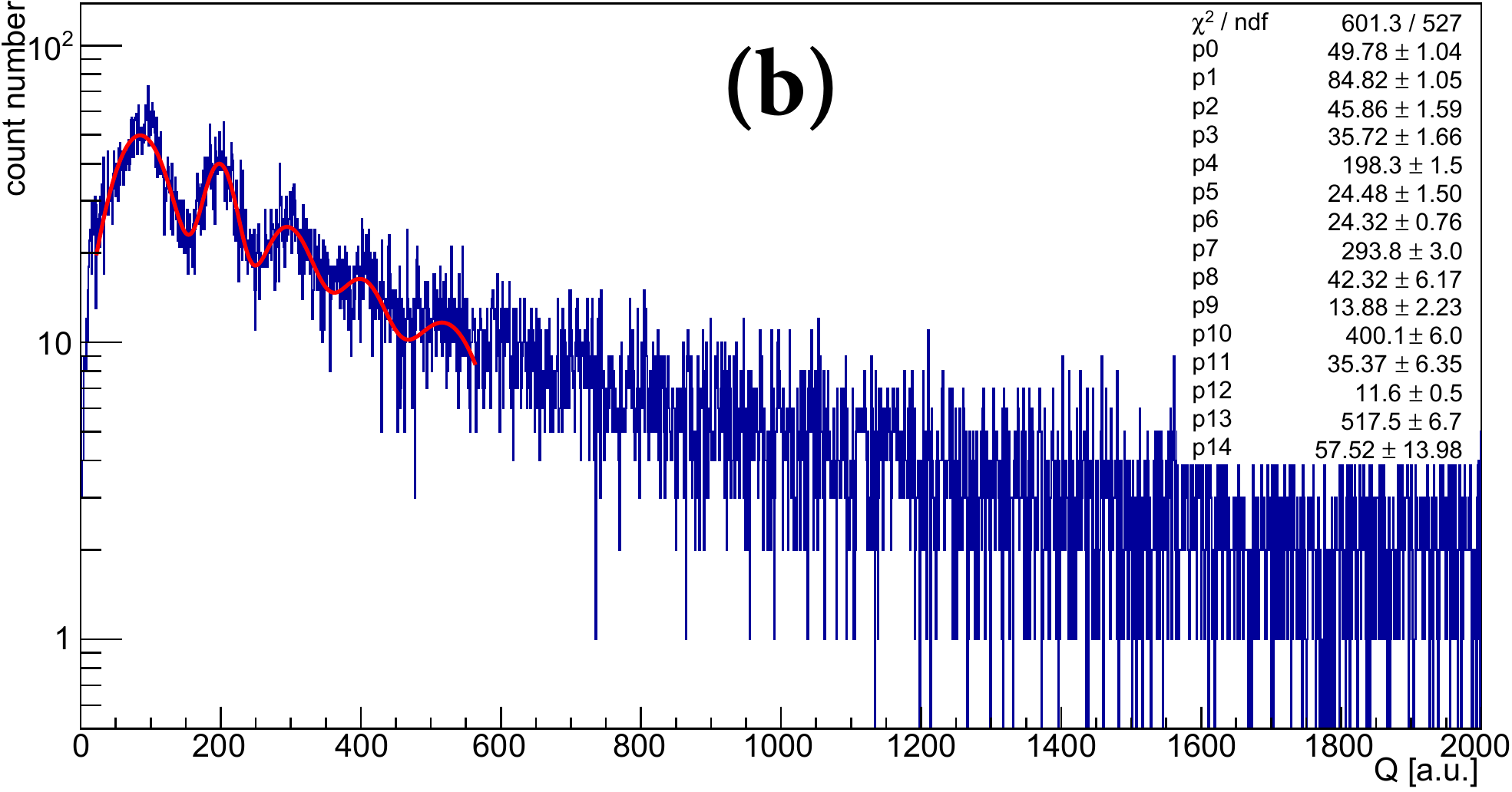}
   
   \vspace{2.5mm}
   
   \includegraphics[width=0.48\linewidth]{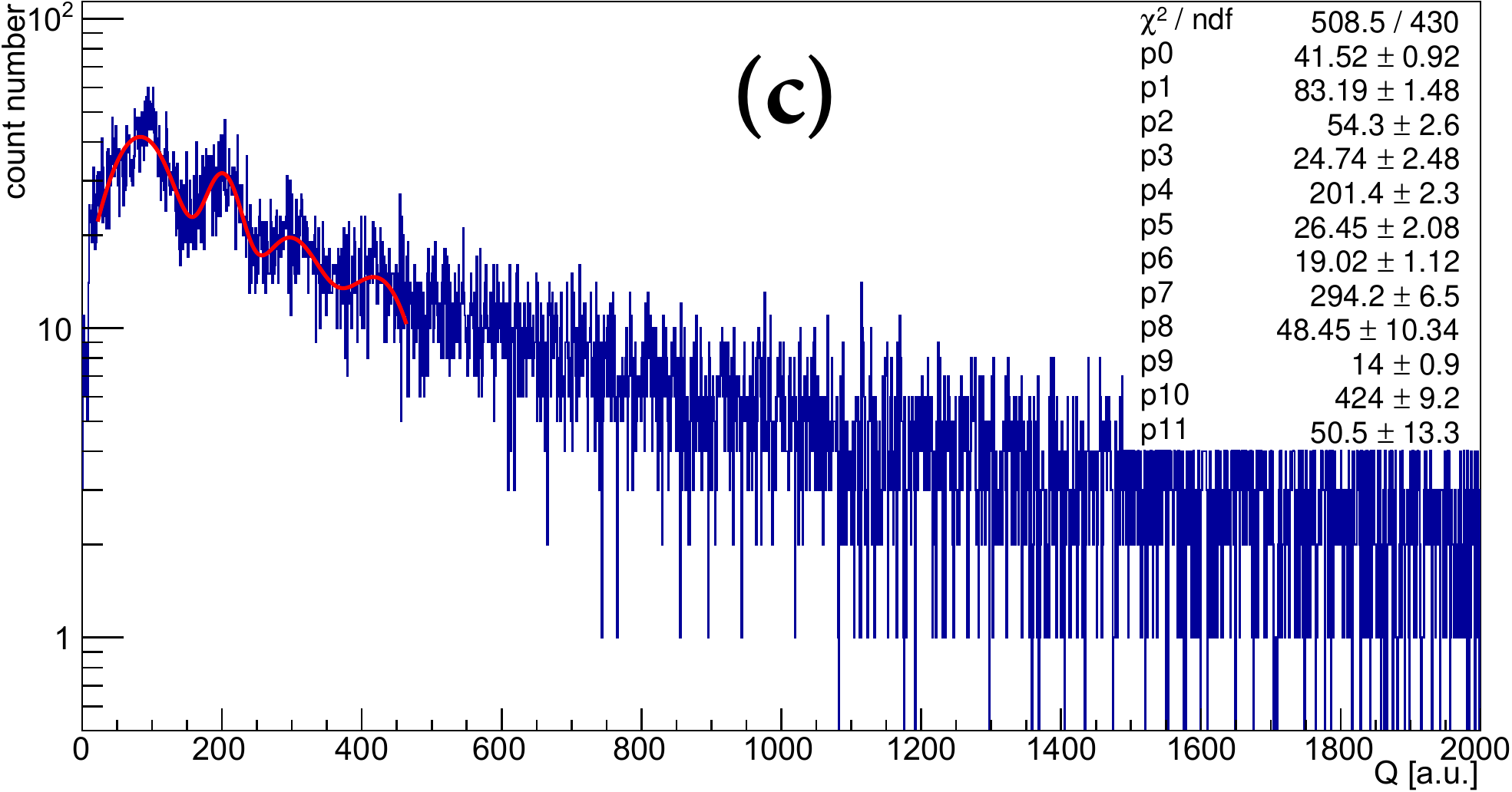} ~~\includegraphics[width=0.48\linewidth]{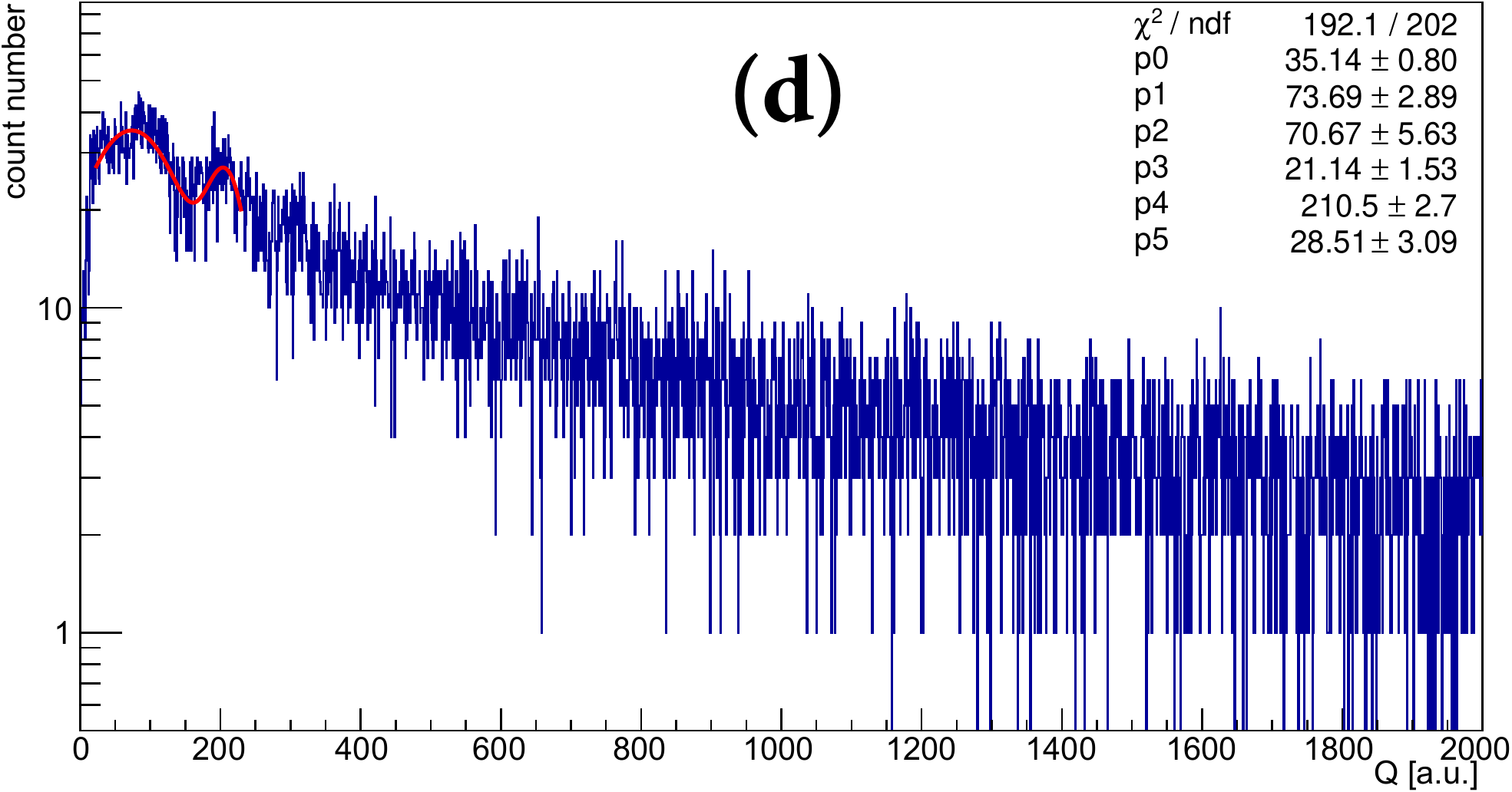}
   
   \vspace{2.5mm}
   
   \includegraphics[width=0.48\linewidth]{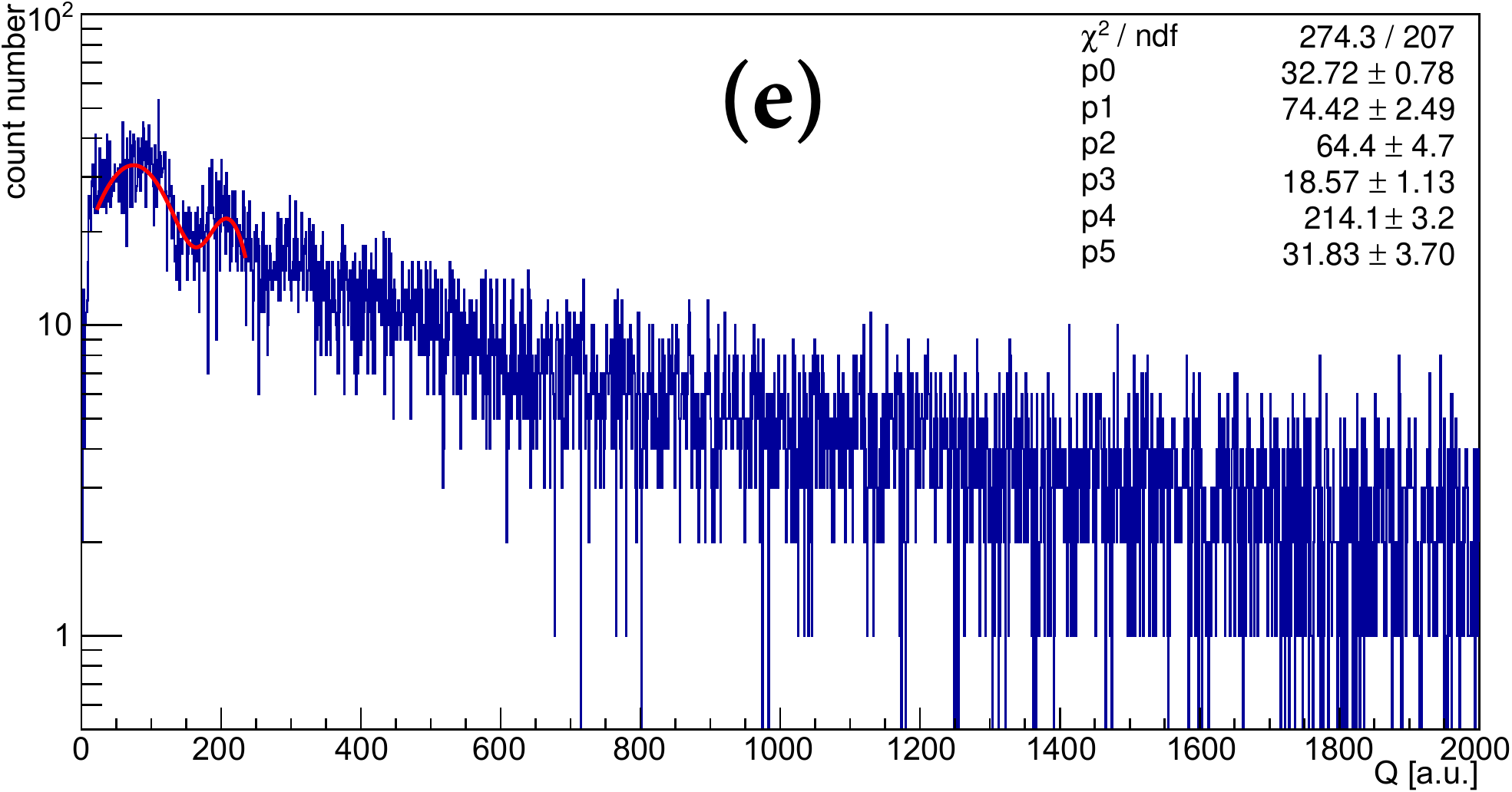} ~~\includegraphics[width=0.48\linewidth]{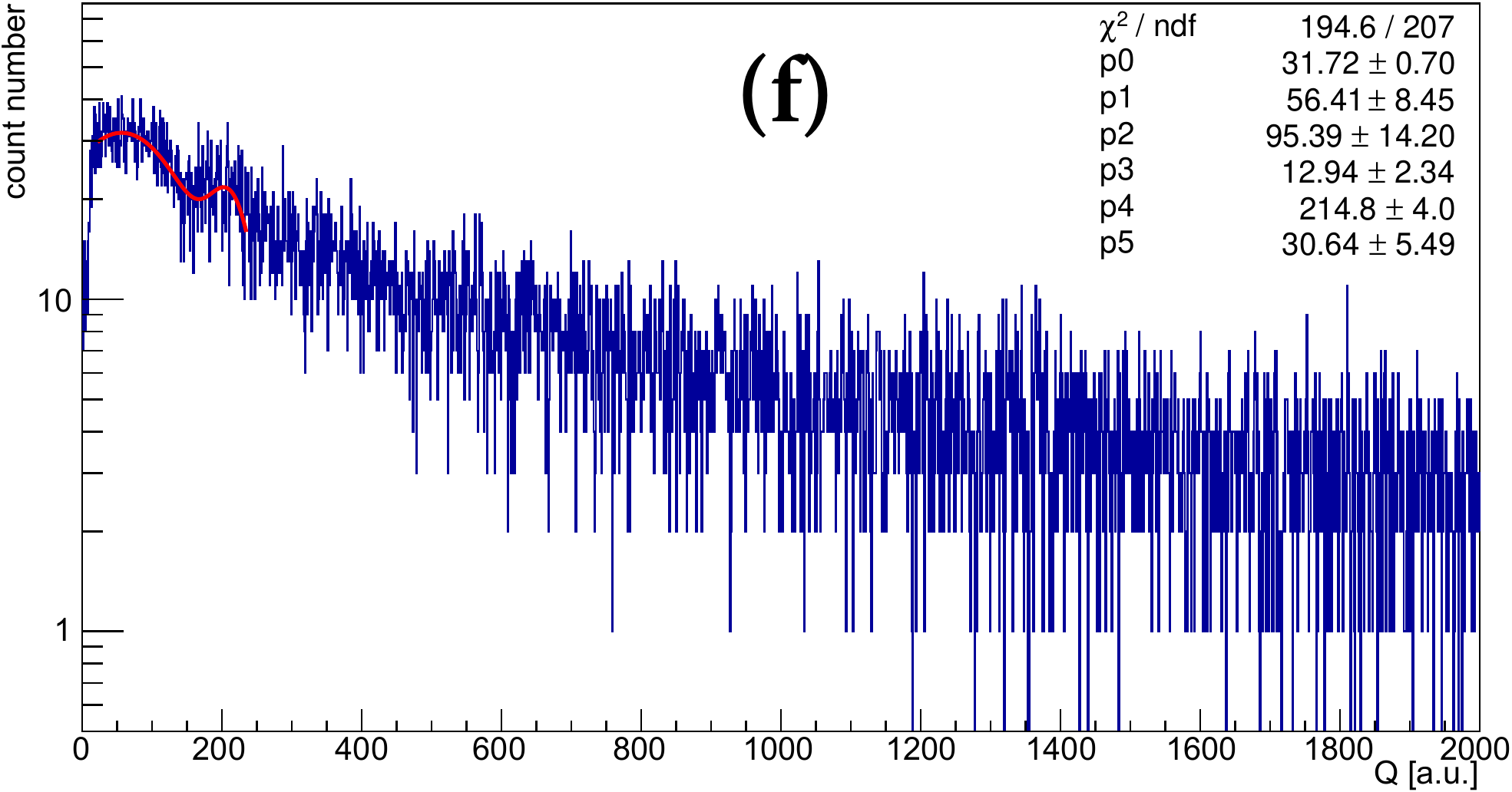}
   \caption{Signal charge distribution, taken in the dark, for the time intervals indicated in the integrated dose plot in Figure~\ref{fig:Doses}. For low radiation levels (distributions (a),(b),(c)) the photoelectron peaks up to 4~p.e. are clearly separated. At higher integrated dose the p.e. peaks start to get smeared.  In the distribution (f) 2 p.e. are barely resolved.}
   \label{fig:QvsD}
\end{figure}

\begin{figure}
    \centering    
 \includegraphics[width=0.6\linewidth]{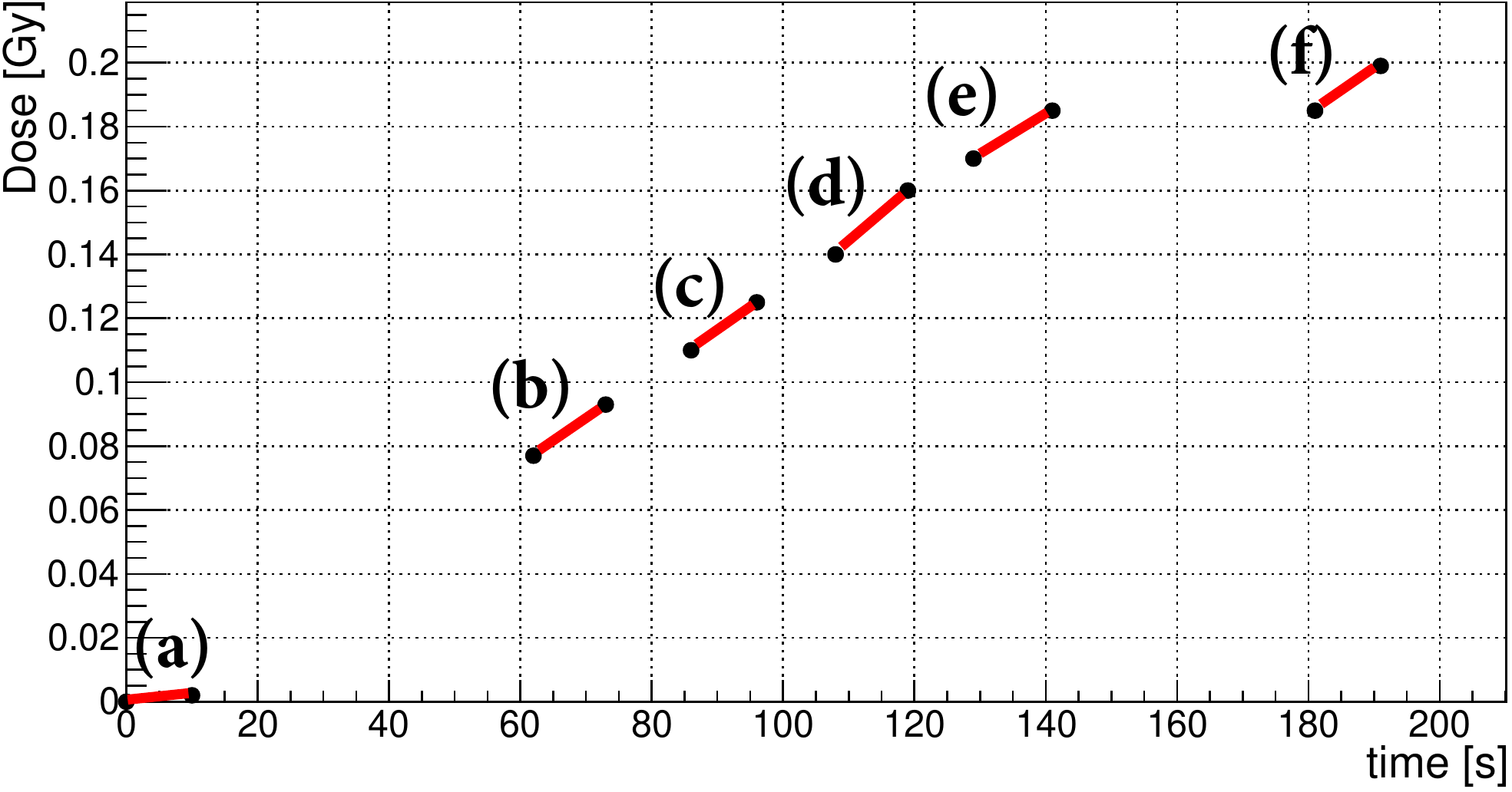}
 \caption{Integrated dose as a function of exposure time.}
\label{fig:Doses}
\end{figure}

The development of the radiation damage can be observed in the signal charge spectra for a short time range with increasing radiation dose. Figure \ref{fig:QvsD} shows the signal charge distributions, taken in the dark, for \SI{10}{\second} time intervals indicated in the integrated dose plot in Figure \ref{fig:Doses}. The damage of the SiPM starts already at rather low integrated dose, of $\sim$~0.2 Gy.

Since this study mostly involves relative measurements, systematic effects that are independent of the radiation exposure cancel. For the gain measurements, the total uncertainty is dominated by the systematic uncertainty related to changes in electronics pickup. To estimate this uncertainty we measured the FWHM of the baseline variations at the different dose values. We found at most $\sim$~\SI{80}{\percent} deviation (at the highest dose value), for widths measured at non-zero radiation dose compared to that in the before exposure. This baseline noise was then added to simulated signal pulses to estimate its effect on the total charge calculation. This Monte Carlo study indicates that the extra noise pickup can lead to systematic errors in the measurement of the gain by the PFA of up to \SI{8}{\percent}. The statistical uncertainty for the gain measurement was found to be negligible. The correlated noise and the dark current rate, on the other hand, are limited by statistical uncertainties.

\subsection{Relative Gain}
\label{sec:RGain}
The gain of a SiPM can be defined as the mean number of output electrons in the single p.e.\ peak. We used the charge distribution of the prompt signal, in Figure~\ref{fig:QvsD}, to study the relative stability in the gain of the SiPM at different integrated dose values, $\it{D}$. The mean value of each individual fitted Gaussian is used to estimate the average charge of the corresponding number of photoelectrons, $\it{Q_{n~p.e.}(D)}$. The slope of the $\it{Q_{n~p.e.}(D)}$ values, when plotted against the number of photoelectrons $\it{n}$, is then used to calculate the average charge of the SiPM single p.e.\ peak response at a specific $\it{D}$, $\bar{Q}(D)$. Thus the stability of the SiPM gain at different radiation doses can be assessed by the ratio, $\eta_{Gain}$, of the charge amplitude, $\it\bar{{Q}}$($\it{D}$) to that in the absence of the exposure:

\begin{equation}
\eta_{Gain} = \frac{\bar{Q}(D)}{\bar{Q}(D=0)}
\end{equation}

Figure~$\ref{fig:SiPMparVsD}$ (I) shows that the relative gain of the AdvanSiD SiPM stays constant as a function of the dose values, with a small deviation of about $\sim$~\SI{10}{\percent} at measurement (f), compared to that before irradiation. However, all variations are consistent with the magnitude of the uncertainties of our measurement.

\subsection{Relative Dark Current Rate}
\label{sec:RDCR}
SiPM dark current signals are mainly caused by thermally generated free charge carriers inside the avalanche region. The pulse of dark noise is similar to that triggered by a photon event. The dark current rate ($\it{DCR}$) is then defined by the rate of the SiPM output pulses, in dark, with amplitude level $\geqslant$ 1~p.e., and can be calculated by the following relation:

\begin{equation}
DCR = \frac{N_{\geqslant 1~ p.e.}}{t_{daq} \cdot A},
\end{equation}
where $\it{N}_{\geqslant 1~ p.e.}$ is the number of the prompt signals with a measured charge of at least 1 p.e., $\it{t}_{daq}$ is the data acquisition time window in seconds, and $\it{A}$ is the surface area of the SiPM.

Figure~$\ref{fig:SiPMparVsD}$ (II) shows $\eta_{DCR}$, the ratio of $\it{DCR}$($\it{D}$) to $\it{DCR}$($\it{D=0}$), as a function of the integrated dose. $\eta_{DCR}$ shows a huge increase, by $\sim$~\SI{30}{\percent} already during the first exposure measurement compared to its value before exposure. Then it stays constant over the irradiation period.

\subsection{Prompt Cross-Talk Probability}
\label{sec:PCT}
Correlated signals are an important source of noise in SiPMs. They are composed of prompt optical cross-talk and delayed after-pulses. The delayed correlated noise probability is discussed in section \ref{sec:PCN}. The origin of prompt cross-talk can be understood as follows: when undergoing an avalanche, carriers near the p-n junction emit photons, due to the scattering of the accelerated electrons. These photons tend to be at near-infrared wavelengths and can travel substantial distances through the device, including to neighboring microcells where they may initiate secondary Geiger avalanches.  As a consequence, a single primary photon may generate signals equivalent to two or more photoelectrons. The prompt cross-talk probability, $\it{P}_{CT}$, depends on over-voltage, $\it{U_\textrm{ov}}$, which is the excess bias beyond the breakdown voltage, device-dependent barriers for photons (trenches), and the size of the microcells. The probability of prompt cross-talk can be calculated as:

\begin{equation}
P_{CT}=\frac{N_{> 1~ p.e.}}{N_{total}},
\end{equation}
where $\it{N}_{> 1~ p.e.}$ is the number of the prompt signals with a measured charge of at least 1.5 p.e., and $\it{N}_{total}$ is the total number of prompt signals above noise. Figure~$\ref{fig:SiPMparVsD}$ (III) shows $\eta_{P_{CT}}$, the ratio of $\it{P}_{CT}$($\it{D}$) to $\it{P}_{CT}$($\it{D=0}$), as a function of the integrated dose. $\eta_{P_{CT}}$ does not show a dependence on the dose values for measurement (a), within the estimated uncertainties, while starting from measurements (b) to (f) it is reduced by $\sim$~\SI{6}{\percent}, compared to the value in the absence of exposure.

\subsection{Delayed Correlated Noise Probability}
\label{sec:PCN}
Both after-pulsing and delayed cross-talk events originate from an existing pulse. After-pulsing is due to the carriers trapped in silicon defects during the avalanche multiplication, then released later during the recharge phase of the microcell. Delayed cross-talk is generated by a similar mechanism to prompt cross-talk. The difference is that the photons generated during the avalanche process are absorbed in the inactive regions of the neighboring cells instead. It takes some time for the minority charge carriers to diffuse into the active region, causing a delayed signal.  In our measurement, we cannot separate after-pulsing from delayed cross-talk and we count them together as delayed correlated noise.

To estimate the delayed correlated noise probability, $\it{P}_{CN}$, we count the number, $N$, of clearly separated pulses occurring immediately after the primary pulse. The time window used for the pulses integration is limited by the acquisition window. The primary pulse time window is found to be $\sim$~\SI{15}{\nano\second}. $\it{P}_{CN}$ is then estimated by normalizing $N$ to the total number of events that contain prompt signals, $N_{prompt}$:

\begin{equation}
P_{CN}=\frac{N}{N_{prompt}}
\end{equation}

\begin{figure}
   \centering
   \includegraphics[width=0.66\linewidth]{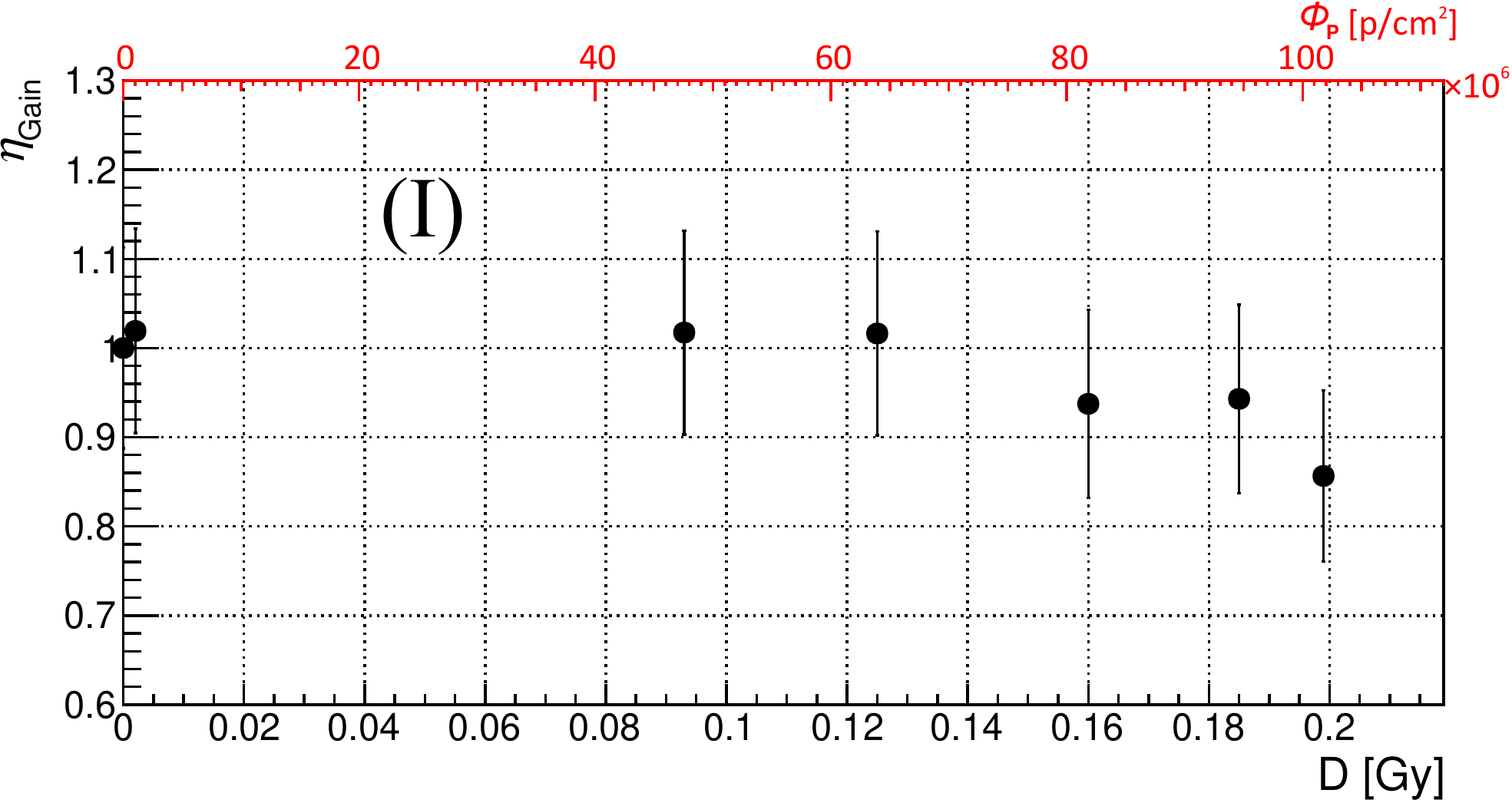}
   \vspace{1.5mm}
   \includegraphics[width=0.66\linewidth]{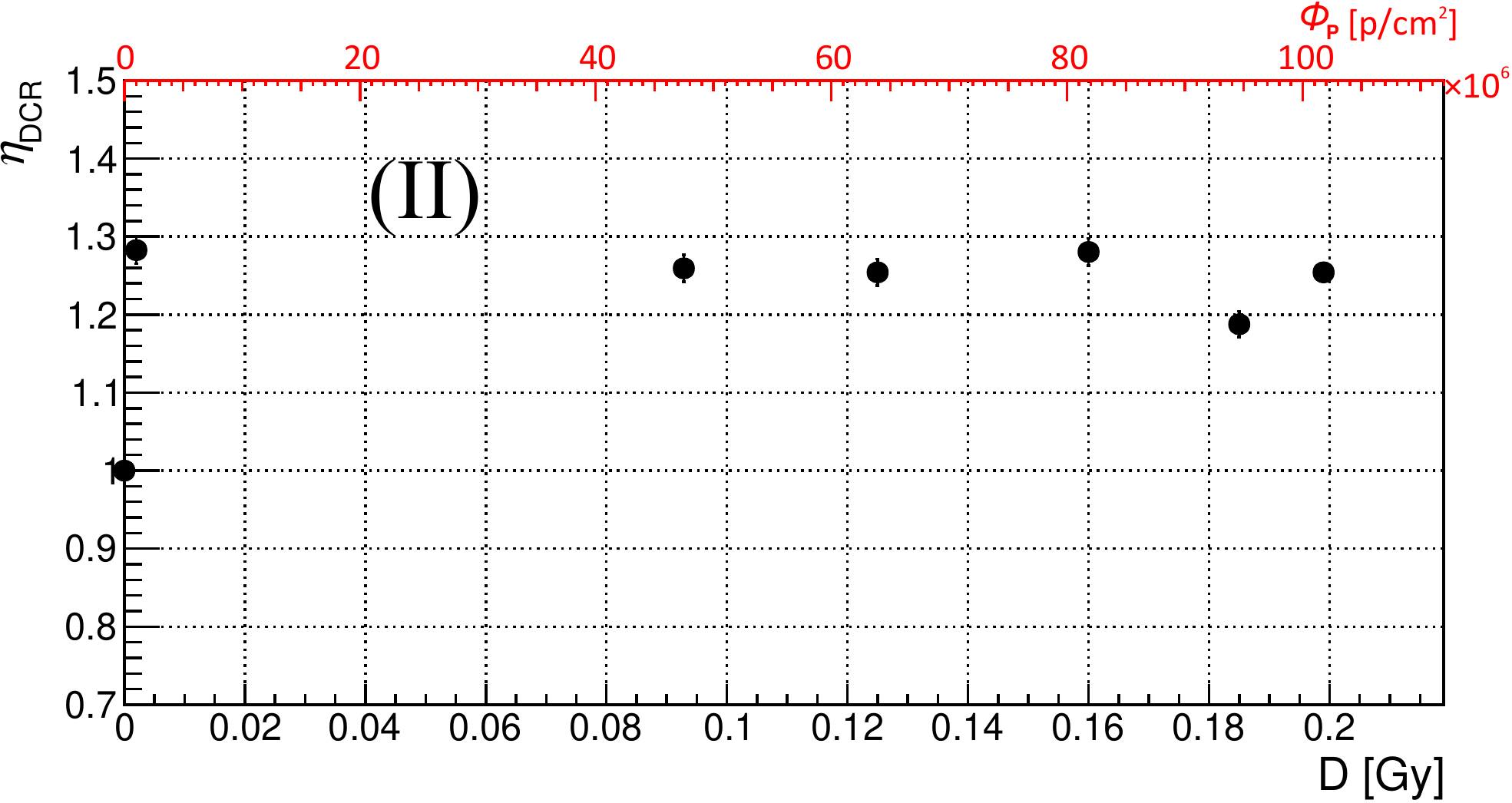}
   \vspace{1.5mm}
   \includegraphics[width=0.66\linewidth]{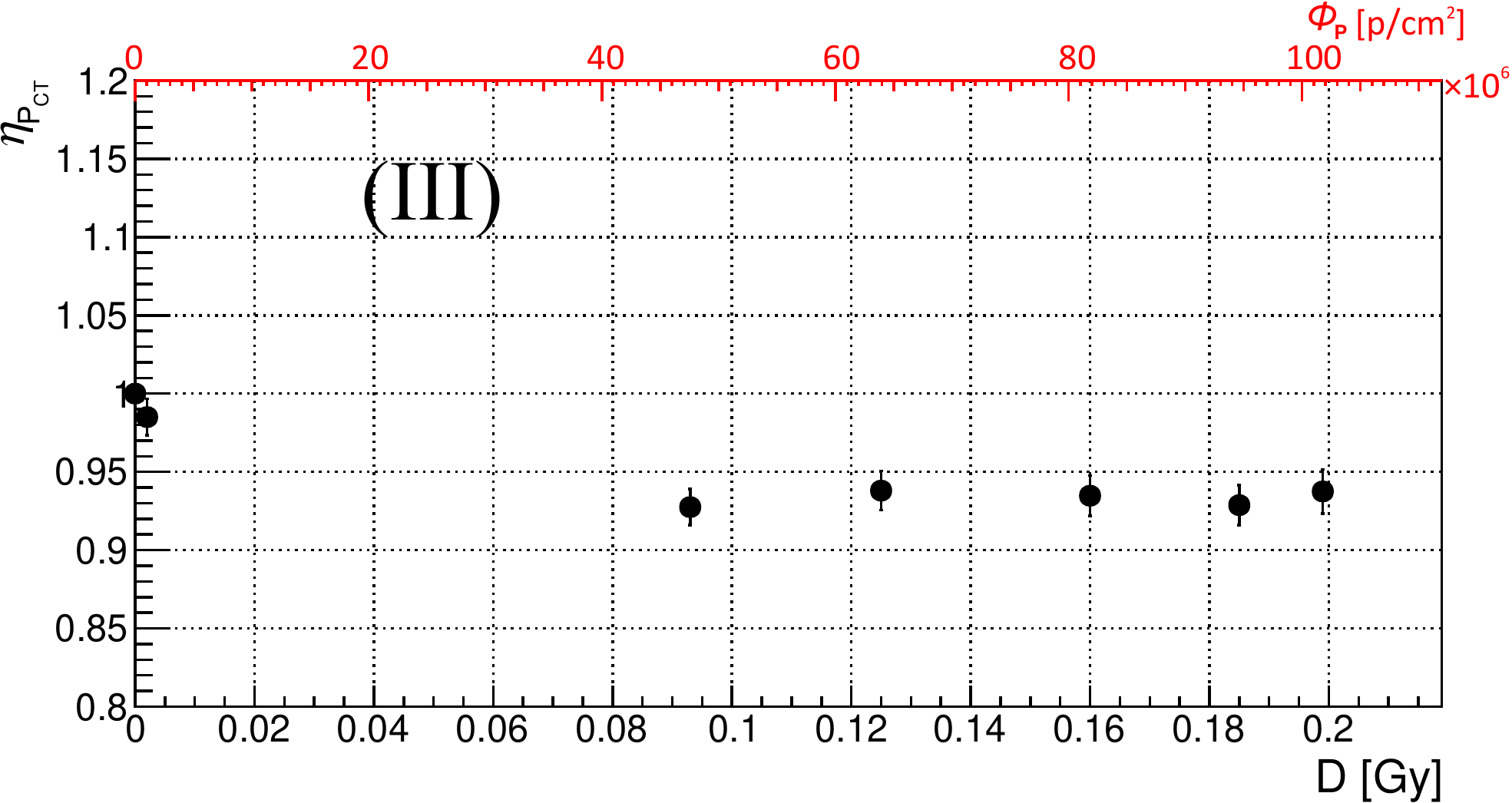}
   \vspace{1.5mm}
   \includegraphics[width=0.66\linewidth]{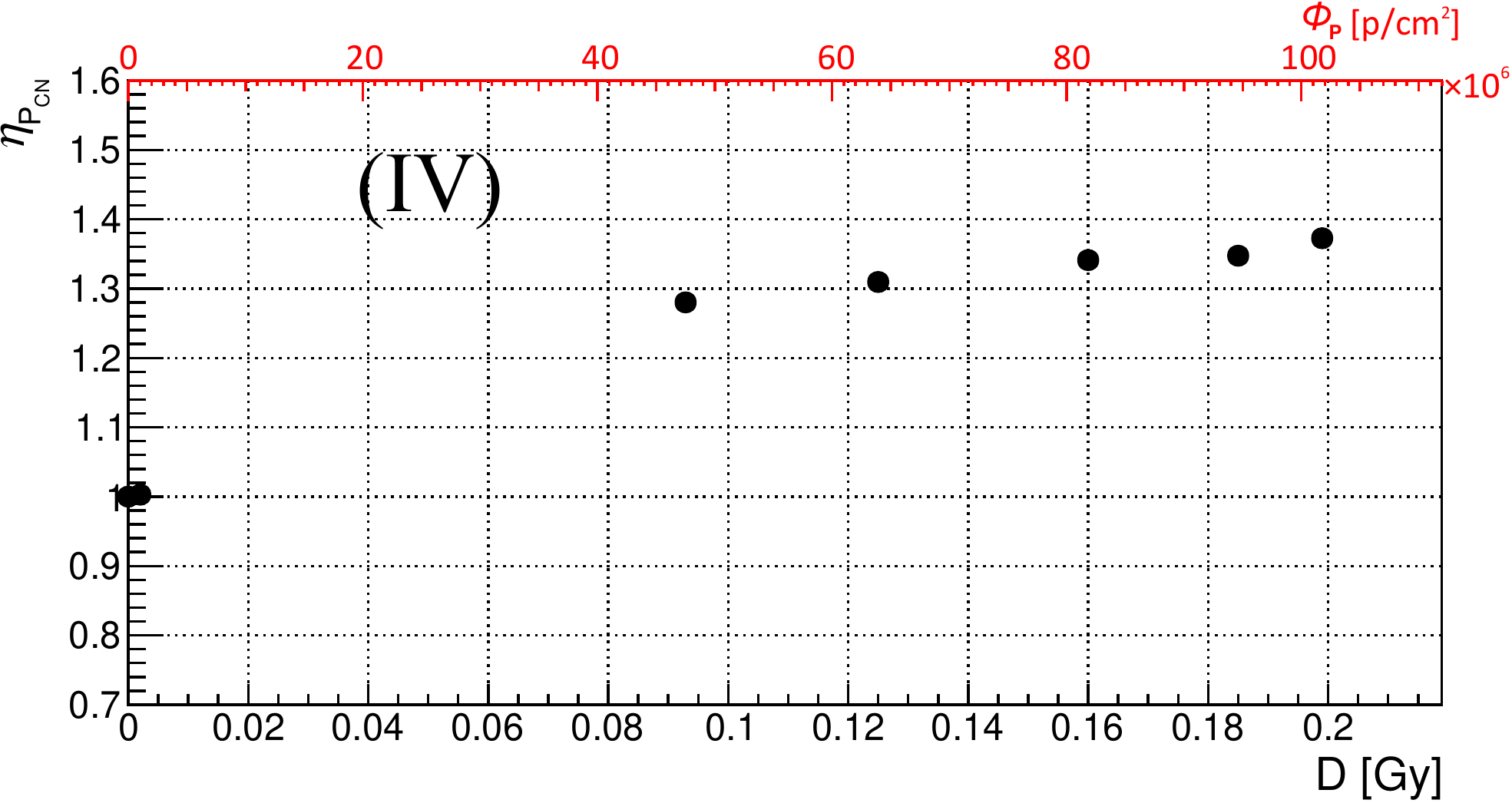}
   \caption{Shown are measurements (a) to (f) during irradiation and one measurement (the first) before irradiation. The relative gain (I), DCR (II), the prompt CT-probability (III) and the delayed correlated noise (IV) are plotted as a function of radiation dose (bottom x-axis) and proton flux (top x-axis).}
   \label{fig:SiPMparVsD}
\end{figure}

Figure~$\ref{fig:SiPMparVsD}$ (IV) shows $\eta_{P_{CN}}$ as a function of the integrated dose. It shows a constant response for the low dose measurement (a), while it increased by $\sim$~\SI{30}{\percent} to $\sim$~\SI{40}{\percent} for measurements (b) to (f), compared to the value in the absence of exposure.

\section{Conclusions}
\label{sec:conclusions}

The radiation hardness of the photosensors is an important characteristic that needs to be carefully studied, especially for those devices that will have high exposure time. Due to the increased use of SiPMs as photosensors the effect of radiation exposure on their performance is a mandatory investigation. Many studies in this field are already available with the general feature of an increased dark current. However, the usability of SiPMs as a function of radiation dose is not well defined due to the manufacturer dependent variation of the detailed structure and the large differences in the detected signals. 

Based on the available data, a rather high tolerable radiation level was expected, but already at a very low integrated dose drastic effects on the output signals were observed. A dose of only $\sim$~0.2 Gy was sufficient to result in a drastic increase of the dark current and a complete dissolution of separate photoelectron peaks. While a generic conclusion on the use of SiPM sensors, as a standalone single photon detectors after the exposure to a certain radiation level is very hard, all devices studied in this work showed a similar insensitivity to the single photon level due to the high dark count rate resulting from the exposure to the radiation level reported here.
But even the exposure to rather high integrated dose doesn't destroy the SiPMs completely. The functionality as photosensor with avalanche behavior at higher light signals is maintained and a triggered readout can separate the signal to be measured from the dark counts.
In all irradiated SiPMs in our analysis we identified a clear reduction of the breakdown voltage by up to about \SI{0.4}{\volt}, resulting in much higher signals and dark count rate  when operating at a fixed operating voltage. This study shows also that this effect should be considered by lowering the voltage to operate the device at a fixed over-voltage rather than a fixed operating voltage, especially for the experiments that depend on detecting light signals with intensities higher than single photoelectrons. A detailed knowledge of the SiPM damage induced by radiation requires more careful studies with devices with a well known internal structure to disentangle the influence of relevant parameters.

These irradiation studies are a first step of SiPM radiation hardness investigations and will be continued with higher proton energies close to the minimum ionizing region, which is more relevant for the typical scintillator readout in most particle detectors. 

\section{Acknowledgments}
\label{sec:Ack}
We thank the crew of the JULIC cyclotron at the Nuclear Physics Institute at Forschungszentrum J\"ulich GmbH for the delivery of the well defined low intensity cyclotron beam and the precise determination of the applied dose rate. This work has been in part funded by the Deutsche Forschungsgemeinschaft (DFG, German Research Foundation) - Projektnummer 423761110.

\bibliography{references}

\end{document}